\documentclass[english,prd,superscriptaddress,nofootinbib,preprintnumbers]{revtex4}
\usepackage[latin1]{inputenc}
\usepackage{amsmath}
\usepackage{amssymb}
\usepackage{graphicx}
\usepackage{esint}
\usepackage{amsthm}
\usepackage{amsthm}\usepackage{bm}
\usepackage{color}

\makeatletter
\usepackage{amsfonts}\usepackage{dcolumn}

\def\be{\begin{equation}}
\def\ee{\end{equation}}
\def\ba{\begin{eqnarray}}
\def\ea{\end{eqnarray}}
\def\bs{\begin{subequations}}
\def\es{\end{subequations}}

\newcommand{\wbd}{\omega_{\rm BD}}

\newcommand{\V}{{\text{\tiny $V$}}}
\newcommand{\Mpl}{M_{\rm pl}}

\@ifundefined{definecolor}{\usepackage{color}}{}

\usepackage{babel}

\makeatother

\usepackage{babel}
\begin{document}

\title{Chaotic inflation in modified gravitational theories}

\author{Antonio De Felice}

\affiliation{Department of Physics, Faculty of Science, Tokyo University of Science,
1-3, Kagurazaka, Shinjuku-ku, Tokyo 162-8601, Japan}

\author{Shinji Tsujikawa}

\affiliation{Department of Physics, Faculty of Science, Tokyo University of Science,
1-3, Kagurazaka, Shinjuku-ku, Tokyo 162-8601, Japan}

\author{Joseph Elliston}

\affiliation{School of Mathematical Sciences,\\
 Queen Mary, University of London, London E1 4NS, UK}

\author{Reza Tavakol}

\affiliation{School of Mathematical Sciences,\\
 Queen Mary, University of London, London E1 4NS, UK}

\begin{abstract}

We study chaotic inflation in the context of modified gravitational
theories. Our analysis covers models based on (i) a field coupling
$\omega(\phi)$ with the kinetic energy
$X=-(1/2)g^{\mu\nu}\partial_{\mu}\phi\partial_{\nu}\phi$ and
a nonmimimal coupling $\zeta\phi^{2}R/2$ with a Ricci scalar $R$,
(ii) Brans-Dicke (BD) theories, (iii) Gauss-Bonnet (GB) gravity, and
(iv) gravity with a Galileon correction. Dilatonic coupling with the kinetic
energy and/or negative nonminimal coupling are shown to lead to
compatibility with observations of the Cosmic Microwave Background
(CMB) temperature anisotropies for the self-coupling inflaton potential
$V(\phi)=\lambda\phi^{4}/4$. BD theory with a quadratic inflaton
potential, which covers Starobinsky's $f(R)$ model $f(R)=R+R^{2}/(6M^{2})$
with the BD parameter $\omega_{{\rm BD}}=0$, gives rise to a smaller
tensor-to-scalar ratio for decreasing $\omega_{{\rm BD}}$. In the
presence of a GB term coupled to the field $\phi$, 
we express the scalar/tensor spectral indices 
$n_{{\rm s}}$ and $n_{\rm t}$ as well as the tensor-to-scalar ratio $r$ 
in terms of two slow-roll parameters and place bounds on 
the strength of the GB coupling from the joint data analysis 
of WMAP 7yr combined with other observations. 
We also study the Galileon-like
self-interaction $\Phi(\phi)X\square\phi$ with exponential coupling
$\Phi(\phi)\propto e^{\mu\phi}$. 
Using a CMB likelihood analysis we put bounds on the strength of the Galileon coupling
and show that the self coupling potential can in fact 
be made compatible with observations in the
presence of the exponential coupling with $\mu>0$.

\end{abstract}

\date{\today}

\pacs{98.80.Cq, 04.60.Pp}

\maketitle

\section{Introduction}

%%%%%%%%%%%%%%%%%%%%%%

Inflation, which was originally proposed by a number of authors independently
in the early 1980s \cite{Star80,infori}, is at present the main theoretical framework
employed in describing the early universe evolution and accounting
for the observational data, specially the observed spectrum of primordial
perturbations \cite{infper}. The sustained success of this framework
over the last 3 decades has been impressive, particularly given the
enormous improvement in the accuracy and resolution of the cosmological
data over this period \cite{COBE,WMAP1,WMAP7,Tegmark,Reid}. The simplest
and most common models of inflation considered so far have employed
a single scalar field, minimally coupled to the curvature and possessing
a canonical kinetic term (see \cite{review} for reviews). As a result,
until recently, much effort has gone into the study of such models,
an important example of which has been the chaotic inflationary model
\cite{Linde-83}.

Despite its successes in accounting for important features of observations,
however, there is no unique mechanism which underpins inflation. Indeed
almost since its inception it has been known that an accelerated phase
of cosmic evolution could be produced by a wide range of mechanisms
such as $f(R)$ theories (see the reviews \cite{Sotiriou,DT10} and references
therein). Thus, an important task in cosmology has been to narrow
down the range of possible alternatives and ultimately to situate
inflationary models within fundamental theories of physical interactions.
There have been two approaches to this problem. The first aims to construct
individual models that are directly suggested by supersymmetric theories.
For example, chaotic inflationary models have been constructed
in the framework of supergravity \cite{supergravity} or superstring
theory \cite{superstring}. The second, on the other hand, considers
classes of generalized models of inflation which possess ingredients
motivated by field theories such as string theory 
\cite{Futamase,Fakir,Bez,GBearly,Armendariz,Ohta,Satoh,Nicolis,Cremi,Kobayashi-etal-10}.

In the absence of a unique, fully successful and non-fine tuned model
of the first kind so far, a great deal of effort has recently gone
into the study of the models of the second kind. In general such models
are expected to possess a number of ingredients motivated by fundamental
theories, including (a) nonminimal couplings of the field to the
Ricci scalar $R$, (b) non-canonical kinetic terms, and (c) higher
derivative quantum gravity corrections in their actions, such as the
Gauss-Bonnet term 
\begin{equation}
{\cal G}\equiv R^{2}-4R_{\alpha\beta}R^{\alpha\beta}
+R_{\alpha\beta\gamma\delta}R^{\alpha\beta\gamma\delta}\,,\label{gb}
\end{equation}
where $R$ is the Ricci scalar, $R_{\alpha\beta}$ 
is the Ricci tensor and $R_{\alpha\beta\gamma\delta}$
is the Riemann tensor, or a nonlinear field-interaction, 
for example in the form 
\begin{equation}
G(\phi,X)\square\phi\,,\label{Gal}
\end{equation}
where $G$ is in general a differential function of the field $\phi$
and $X=-(1/2)g^{\mu\nu}\partial_{\mu}\phi\partial_{\nu}\phi$. 
In addition, such models would also
be expected to possess multiple scalar fields, but in order to
separate the effects of different ingredients we shall confine ourselves
to models possessing a single scalar field.

In standard chaotic inflation the self coupling $\lambda$ for the
inflaton potential $V(\phi)=\lambda\phi^{4}/4$ is constrained to
be small ($\lambda \approx 10^{-13}$) from the WMAP normalization \cite{WMAP1}.
Moreover, this model is in tension with the observations of CMB temperature
anisotropies because it predicts a large tensor-to-scalar ratio ($r \approx 0.3$)
\cite{WMAP7}. If we take into account the nonminimal coupling $\zeta\phi^{2}R/2$,
it is possible to realise larger $\lambda$ compatible with the natural
values appearing in particle physics ($\lambda=0.01$-0.1). In the
limit where the negative nonminimal coupling $\zeta$ satisfies the condition 
$|\zeta|\gg1$, the tensor-to-scalar ratio $r$ reduces to the order of $10^{-3}$
with the scalar spectral index $n_{{\rm s}}\approx0.96$ \cite{nonminimalper}.
This agrees well with the CMB observations \cite{Komatsuper,Gumjudpai,Linde}.
Recently, there has been renewed interest in nonminimally coupled
inflation models by identifying the inflaton as a Higgs boson appearing
in standard model of particle physics \cite{Bez,Higgspapers}.

It is of interest to see whether the chaotic inflationary models 
that are in tension with observations can be rescued 
by taking into account the various field couplings mentioned
above. In the presence of the nonlinear kinetic interaction $(1/M^{3})X\square\phi$
that respects the Galilean symmetry $\partial_{\mu}\phi\to\partial_{\mu}\phi+b_{\mu}$
in Minkowski spacetime, for example, it was recently shown that
the tensor-to-scalar ratio can reduce to $r\simeq0.18$ for the potential
$V(\phi)=\lambda\phi^{4}/4$ \cite{Kamada-etal-10}. Moreover, the
coupling constant $\lambda$ is related to the mass scale $M$ by the
WMAP normalization, which allows for the possibility  
of having a natural coupling of the order of $\lambda=0.01$-0.1. 
Of course this outcome would be expected to be different 
if we choose more general functions of
$G(\phi,X)$ that depend on both $\phi$ and $X$.

In this paper we wish to make a detailed and unified study of inflation
in the context of modified gravitational theories. To this end, and
for concreteness and comparison with previous works, we shall study
chaotic inflation, sourced by potentials of the form 
\begin{equation}
V(\phi)=V_{0}(\phi/\Mpl)^{p}\quad , \quad(p>0)\,,
\label{chaoticpo}
\end{equation}
where $V_{0}$ and $p$ are real constants and 
$\Mpl=1/\sqrt{8\pi G_{N}}=2.44\times10^{18}$~GeV
is the reduced Planck mass ($G_{N}$ is the gravitational constant). We
shall consider a number of field couplings such as (i) the non-canonical kinetic term 
$\omega(\phi)X$ as well as the nonmimimal
coupling $\zeta\phi^{2}R/2$, (ii) Brans-Dicke (BD) theories having
explicit couplings $\phi R$ and $(\omega_{{\rm BD}}/\phi)X$ (including
$f(R)$ gravity), (iii) Gauss-Bonnet (GB) coupling of the form $\xi_0 e^{\mu\phi/\Mpl}{\cal G}$,
and (iv) the generalized Galileon coupling of the form 
$(e^{\mu\phi/\Mpl}/M^{4n-1})X^n \square\phi$
(which reproduces the pure Galileon term \cite{Nicolis,Galipapers} in the limit that $\mu \to 0$ and 
$n \to 1$). The terms of the forms (i), (iii), and (iv) appear
as a next order correction to the tree-level action in low energy
effective string theory \cite{corrections}.

In each model we evaluate the three inflationary observables: (a)
the scalar spectral index $n_{{\rm s}}$, (b) the tensor spectral
index $n_{{\rm t}}$, and (c) the tensor-to-scalar ratio $r$. We
place observational constraints on the model parameters by carrying
out a CMB likelihood analysis. We find that in most cases it is
possible for the chaotic inflationary potentials with $p=2$ and $p=4$
to be consistent with the current observations. We also show that
the equilateral nonlinear parameter $f_{{\rm NL}}^{{\rm equil}}$
describing the scalar non-Gaussianity is smaller than the order of
unity, apart from in the generalized Galileon model with $n \gg 1$.

The structure of the paper is as follows. In Sec.~\ref{modelsec}
we derive the background equations of motion for the general action
(\ref{action}) and introduce a number of slow-roll parameters. In
Sec.~\ref{cperturbations} we present the power spectra of scalar
and tensor perturbations derived under the framework of linear cosmological
perturbation theory. The formula of the equilateral non-Gaussianity
parameter $f_{{\rm NL}}^{{\rm equil}}$ is also given there. In Sec.~\ref{Einsec}
we show the three inflationary observables $n_{{\rm s}}$, $n_{{\rm t}}$,
and $r$ in the Einstein frame for the theories without the GB or
Galileon terms, which are convenient for the analysis in Secs.~\ref{nonminimalsec}
and \ref{BDsec}. Sec.~\ref{nonminimalsec} is devoted to the study
of the nonmimimal coupling $\zeta\phi^{2}R/2$ as well as the coupling
$e^{\mu\phi/\Mpl}X$. In Sec.~\ref{BDsec} we consider BD theories
with the two potentials $V=V_{0}(\phi/\Mpl)^{p}$ and $V(\phi)=V_{0}(\phi-\Mpl)^{p}$
in the Jordan frame for arbitrary BD parameters $\omega_{{\rm BD}}$.
For $p=2$ and $\omega_{{\rm BD}}=0$ the latter potential covers
the Starobinsky's $f(R)$ model $f(R)=R+R^{2}/(6M^{2})$. In Sec.~\ref{GBsec}
we study chaotic inflation in the presence of the exponential GB coupling
and place observational constraints on the strength of the GB coupling.
In Sec.~\ref{Galileonsec} we show that the exponential Galileon
coupling can lead to the consistency of self-coupling 
chaotic inflation with the observational data.
Sec.~\ref{conclusions} is devoted to our conclusions.

%%%%%%%%%%%%%%%%%%%%%%%%%%%%%%%%%%%%%%%%%%%%%
\section{Models and background equations}
\label{modelsec} 
%%%%%%%%%%%%%%%%%%%%%%%%%%%%%%%%%%%%%%%%%%%%

We start with the generalized action 
\begin{equation}
S=\int d^{4}x\sqrt{-g}\left[\frac{M_{{\rm pl}}^{2}}{2}F(\phi)R
+\omega(\phi)X-V(\phi)-\xi(\phi){\cal G}-G(\phi,X)\square\phi\right]\,,
\label{action}
\end{equation}
where $g$ is a determinant of the space-time metric $g_{\mu\nu}$,
and $\phi$ is a scalar field with a kinetic term $X$. 
The functions $F(\phi)$, $\omega(\phi)$, and $\xi(\phi)$ are differentiable
functions of $\phi$, whereas $G(\phi,X)$ depend on both $\phi$
and $X$. The field $\phi$ couples to both the Ricci scalar
$R$ as well as the Gauss-Bonnet term ${\cal G}$.

We consider the flat Friedmann-Lema\^{i}tre-Robertson-Walker (FLRW)
space-time with a scale factor $a(t)$, where $t$ is cosmic time.
The background equations are then given by 
\begin{eqnarray}
E_{1} & \equiv & 3\Mpl^{2}FH^{2}+3\Mpl^{2}H\dot{F}-\omega X-V-24H^{3}\dot{\xi}-6H\dot{\phi}XG_{,X}+2XG_{,\phi}=0\,,\label{E1eq}\\
E_{2} & \equiv & 3\Mpl^{2}FH^{2}+2\Mpl^{2}H\dot{F}+2\Mpl^{2}F\dot{H}+\Mpl^{2}\ddot{F}+\omega X-V-16H^{3}\dot{\xi}-16H\dot{H}\dot{\xi}-8H^{2}\ddot{\xi}-G_{,X}\dot{\phi}\dot{X}-G_{,\phi}\dot{\phi}^{2}=0,\label{E2eq}\\
E_{3} & \equiv & (\omega+6H\dot{\phi}G_{,X}+6H\dot{\phi}\, XG_{,XX}-2XG_{,\phi X}-2G_{,\phi})\ddot{\phi}\nonumber \\
 &  & +(3\omega H+\dot{\phi}\,\omega_{,\phi}+9H^{2}\dot{\phi}G_{,X}+3\dot{H}\dot{\phi}G_{,X}+3H\dot{\phi}^{2}G_{,\phi X}-6HG_{,\phi}-G_{,\phi\phi}\dot{\phi})\dot{\phi}\nonumber \\
 &  & -\omega_{,\phi}X+V_{,\phi}-6\Mpl^{2}H^{2}F_{,\phi}-3\Mpl^{2}\dot{H}F_{,\phi}+24H^{4}\xi_{,\phi}+24H^{2}\dot{H}\xi_{,\phi}=0\,,\label{E3eq}
\end{eqnarray}
where $H\equiv\dot{a}/a$ is the Hubble parameter, a dot represents
a derivative with respect to $t$, and a comma represents a partial derivative 
in terms of $\phi$ or $X$.
Only two of the above equations are independent due to the Bianchi identities, 
$\dot{\phi}E_{3}+\dot{E}_{1}+3H(E_{1}-E_{2})=0$.
The combined equation, $(E_{2}-E_{1})/(\Mpl^{2}H^{2}F)=0$, gives 
\begin{equation}
\epsilon\equiv-\frac{\dot{H}}{H^{2}}=-\frac{\dot{F}}{2HF}
+\frac{\ddot{F}}{2H^{2}F}+\frac{\omega X}{\Mpl^{2}H^{2}F}
+\frac{4H\dot{\xi}}{\Mpl^{2}F}-\frac{8\dot{H}\dot{\xi}}{\Mpl^{2}HF}
-\frac{4\ddot{\xi}}{\Mpl^{2}F}+\frac{3\dot{\phi}XG_{,X}}{\Mpl^{2}HF}
-\frac{\ddot{\phi}XG_{,X}}{\Mpl^{2}H^{2}F}
-\frac{2XG_{,\phi}}{\Mpl^{2}H^{2}F}\,.\label{dotHeq}
\end{equation}
Since $\epsilon\ll1$ during inflation, the modulus of each term on the r.h.s. of
Eq.~(\ref{dotHeq}) is much smaller than unity (unless some cancellation
occurs between those terms). We introduce the following slow-roll parameters
\begin{eqnarray}
 &  & \delta_{F}\equiv\frac{\dot{F}}{HF}\,,\quad\delta_{X}\equiv\frac{\omega X}{\Mpl^{2}H^{2}F}\,,\quad\delta_{\xi}\equiv\frac{H\dot{\xi}}{\Mpl^{2}F}\,,\quad\delta_{GX}\equiv\frac{\dot{\phi}XG_{,X}}{\Mpl^{2}HF}\,,\quad\delta_{\phi}\equiv\frac{\ddot{\phi}}{H\dot{\phi}}\,,\quad\delta_{G\phi}\equiv\frac{XG_{,\phi}}{\Mpl^{2}H^{2}F}\,,\nonumber \\
 &  & \eta_{F}\equiv\frac{\dot{\delta}_{F}}{H\delta_{F}}\,,\quad\eta_{\xi}\equiv\frac{\dot{\delta}_{\xi}}{H\delta_{\xi}}\,,\label{slowvariation}
\end{eqnarray}
by which we have 
\begin{equation}
\frac{\ddot{F}}{H^{2}F}=\delta_{F}(\delta_{F}+\eta_{F}-\epsilon)\,,\qquad\frac{\ddot{\xi}}{\Mpl^{2}F}=\delta_{\xi}(\delta_{F}+\eta_{\xi}+\epsilon)\,.
\end{equation}
{}From Eq.~(\ref{dotHeq}) we obtain 
\begin{eqnarray}
\epsilon & = & \frac{2\delta_{X}-\delta_{F}+8\delta_{\xi}+6\delta_{GX}-4\delta_{G\phi}+\delta_{F}(\delta_{F}+\eta_{F})-8\delta_{\xi}(\delta_{F}+\eta_{\xi})-2\delta_{\phi}\delta_{GX}}{2+\delta_{F}-8\delta_{\xi}}\\
 & = & \delta_{X}-\delta_{F}/2+4\delta_{\xi}+3\delta_{GX}
 -2\delta_{G\phi}+{\cal O}(\epsilon^{2})\,,\label{epap}
\end{eqnarray}
where in the latter step we have taken the leading-order contribution.

%%%%%%%%%%%%%%%%%%%%%%%%%%%%%%%%%
\section{Cosmological perturbations}
\label{cperturbations} 
%%%%%%%%%%%%%%%%%%%%%%%%%%%%%%%%%

Let us consider cosmological perturbations about the flat FLRW background.
We take into account, up to a gauge choice, both the perturbations 
in the scalar field $\delta \phi$ and in the scalar and tensor modes of the metric.
For the calculations of observables, including primordial non-Gaussianities,
it is convenient to employ the 4-dimensional ADM perturbed metric \cite{ADM} 
of the form
\begin{equation}
ds^{2}=-\left[(1+\alpha)^{2}-a^{-2}(t)e^{-2{\cal R}}
(\partial\psi)^{2}\right]\, 
dt^{2}+2\partial_{i}\psi\, dt\, dx^{i}+a^{2}(t) 
\left( e^{2{\cal R}} \delta_{ij}+h_{ij} \right)
dx^i dx^j\,,
\label{eq:metrica}
\end{equation}
where ${\cal R}$ is the curvature perturbation, $\alpha$ and $\psi$
are related with the lapse 
$(1+\alpha)$ and the shift vector $\partial_{i}\psi$, and
$h_{ij}$ are tensor perturbations.
In the metric (\ref{eq:metrica}) we have gauged away a field
$E$ appearing as $E_{,ij}$ inside $h_{ij}$, to fix
the spatial components of a gauge-transformation vector
$\xi^{\mu}$. We choose the uniform-field gauge where 
$\delta \phi=0$, in order to fix the time-component 
of $\xi^{\mu}$ \cite{Maldacena}.

Expanding the action (\ref{action}) up to second order for the metric
(\ref{eq:metrica}), performing
integration by parts and using the Hamiltonian and momentum constraints
to eliminate the contribution coming from $\alpha$ and $\psi$,
we obtain the following second-order action \cite{DT11} 
\begin{equation}
S_{2}=\int dt\, d^{3}x\, a^{3}Q\left[\dot{{\cal R}}^{2}
-\frac{c_{s}^{2}}{a^{2}}\,(\partial{\cal R})^{2}\right]\,,\label{eq:az2}
\end{equation}
where 
\begin{eqnarray}
Q & \equiv & \frac{w_{1}(4w_{1}w_{3}+9w_{2}^{2})}{3w_{2}^{2}}\,,
\label{eq:defQ}\\
c_{s}^{2} & \equiv & \frac{3(2w_{1}^{2}w_{2}H-w_{2}^{2}w_{4}
+4w_{1}\dot{w}_{1}w_{2}-2w_{1}^{2}\dot{w}_{2})}
{w_1(4w_{1}w_{3}+9w_{2}^{2})}\,,\label{eq:defc2s}
\end{eqnarray}
and 
\begin{eqnarray}
w_{1} & \equiv & \Mpl^{2}\, F-8H\,\dot{\xi}\,,\\
w_{2} & \equiv & \Mpl^{2}(2HF+\dot{F})-2\dot{\phi}XG_{,X}-24H^{2}\dot{\xi}\,,\\
w_{3} & \equiv & -9\Mpl^{2}F{H}^{2}-9\Mpl^{2}H\dot{F}+3\omega X+144H^{3}\dot{\xi}+18H\dot{\phi}(2XG_{,X}+X^{2}G_{,XX})-6(XG_{,\phi}+X^{2}G_{,\phi X})\,,\\
w_{4} & \equiv & \Mpl^{2}F-8\ddot{\xi}\,.
\end{eqnarray}
In order to avoid the appearance of ghosts and Laplacian instabilities 
we require that
\begin{equation}
Q>0\,,\qquad c_{s}^{2}>0\,,
\end{equation}
respectively. One can express $w_{i}$ ($i=1,\cdots,4$) in terms
of the slow-roll parameters. For example one has 
\begin{equation}
w_{3}=-9\Mpl^{2}FH^{2}\left(1+\delta_{F}-\frac{1}{3}\delta_{X}-16\delta_{\xi}-4\delta_{GX}+\frac{2}{3}\delta_{G\phi}-2\delta_{GX}\lambda_{GX}+\frac{2}{3}\delta_{G\phi}\lambda_{G\phi}\right)\,,
\end{equation}
where 
\begin{equation}
\lambda_{GX} \equiv \frac{XG_{,XX}}{G_{,X}}\,,\qquad
\lambda_{G\phi} \equiv\frac{XG_{,\phi X}}{G_{,\phi}}\,.
\end{equation}
The quantities $\lambda_{GX}$ and $\lambda_{G\phi}$
are not necessarily small.

The expansion in terms of the slow-roll parameters gives 
\begin{eqnarray}
 &  & c_{s}^{2}\simeq\frac{\delta_{X}+4\delta_{GX}-2\delta_{G\phi}+2\delta_{G\phi}\lambda_{G\phi}}{\delta_{X}+6\delta_{GX}-2\delta_{G\phi}+6\delta_{GX}\lambda_{GX}-2\delta_{G\phi}\lambda_{G\phi}}\,,\label{cs}\\
 &  & \epsilon_{{\rm s}}\equiv\frac{Qc_{s}^{2}}{\Mpl^{2}F}=\delta_{X}+4\delta_{GX}-2\delta_{G\phi}+2\delta_{G\phi}\lambda_{G\phi}\nonumber \\
 &  & ~~~~~~~~~~~~~~~~~-2\delta_{G\phi}\delta_{F}\lambda_{G\phi}+16\delta_{G\phi}\delta_{\xi}\lambda_{G\phi}+2\delta_{\phi}\delta_{GX}\lambda_{GX}+4\delta_{G\phi}\delta_{GX}\lambda_{G\phi}+3\delta_{F}^{2}/4-12\delta_{\xi}\delta_{F}+2\delta_{GX}\delta_{\phi}-\delta_{F}\delta_{X}\nonumber \\
 &  & ~~~~~~~~~~~~~~~~~-5\delta_{F}\delta_{GX}+2\delta_{F}\delta_{G\phi}+8\delta_{\xi}\delta_{X}+40\delta_{\xi}\delta_{GX}-16\delta_{\xi}\delta_{G\phi}-4\delta_{GX}\delta_{G\phi}+2\delta_{GX}\delta_{X}+48\delta_{\xi}^{2}+7\delta_{GX}^{2}\nonumber \\
 &  & ~~~~~~~~~~~~~~~~~+{\cal O}(\epsilon^{3})\,,\label{eps}
\end{eqnarray}
where in the expression for $c_{s}^{2}$ we have picked up the leading-order
contributions. In standard slow-roll inflation with $F=1$, $\omega=1$, $\xi=0$,
and $G=0$ we obtain the exact expressions $c_{s}^{2}=1$ and $Q/\Mpl^2=\delta_{X}=\epsilon$. 
Equation (\ref{cs}) shows that the nonminimal coupling $F(\phi)R$ and the Gauss-Bonnet
term $\xi(\phi) {\cal G}$ do not give rise to contributions to $c_{\rm s}^{2}$
at linear order. The effects of those terms on $c_{s}^{2}$ appear
at the next order.

The power spectrum of the curvature perturbation is given by \cite{DT11}
\begin{equation}
{\cal P}_{{\rm s}}=\frac{H^{2}}{8\pi^{2}Qc_{s}^{3}}
=\frac{H^{2}}{8\pi^{2}\Mpl^{2}F\epsilon_{s}c_{s}}\,,
\end{equation}
which gives the scalar spectral index 
\begin{eqnarray}
n_{{\rm s}}-1\equiv\frac{d\ln{\cal P}_{{\rm s}}}{d\ln k}
\bigg|_{c_{s}k=aH} 
& = & -2\epsilon-\delta_{Q}-3s\\
& = & -2\epsilon-\delta_{F}-\eta_{s}-s\,,\label{nR}
\end{eqnarray}
where 
\begin{equation}
\delta_{Q}\equiv\frac{\dot{Q}}{HQ}\,,\qquad s\equiv\frac{\dot{c}_{s}}{Hc_{s}}\,,
\qquad\eta_{s}\equiv\frac{\dot{\epsilon}_{s}}{H\epsilon_{s}}\,.\label{etas}
\end{equation}
We have assumed that both $H$ and $c_{s}$ vary slowly, such that
$d\ln k$ at $c_{s}k=aH$ may be approximated by $d\ln k=d\ln a=Hdt$.

The tensor power spectrum is given by \cite{DT11} 
\begin{equation}
{\cal P}_{{\rm t}}=\frac{H^{2}}{2\pi^{2}Q_{t}c_{t}^{3}}\,,
\end{equation}
where $Q_{t}=w_{1}/4=\Mpl^2 F (1-8\delta_{\xi})/4$ and 
$c_{t}^{2}=w_{4}/w_{1}=1+8 \delta_{\xi}+{\cal O} (\epsilon^2)$. 
Taking the leading-order contribution in ${\cal P}_{{\rm t}}$, 
it follows that ${\cal P}_{{\rm t}} \simeq 2H^2/(\pi^2 \Mpl^2 F)$.
The tensor spectral index is
\begin{eqnarray}
n_{{\rm t}}\equiv\frac{d\ln{\cal P}_{{\rm t}}}{d\ln k}\bigg|_{c_t k=aH} 
& = & -2\epsilon-\delta_{F}\,,\label{nT}
\end{eqnarray}
which is valid at first order in slow-roll. At times before the end
of inflation ($\epsilon\ll1$) when both ${\cal P}_{{\rm s}}$ and
${\cal P}_{{\rm t}}$ remain approximately constants, 
we can estimate the tensor-to-scalar ratio, as 
\begin{eqnarray}
r\equiv\frac{{\cal P}_{{\rm t}}}{{\cal P}_{{\rm s}}} \simeq
16\frac{Qc_{s}^{3}}{\Mpl^{2}F}=16c_{s}\epsilon_{s}\,.\label{rgene}
\end{eqnarray}

The non-Gaussianities of scalar perturbations for the action (\ref{action}) 
have been evaluated in Ref.~\cite{DT11} 
(see also Refs.~\cite{Maldacena,Mizuno,Kobayashi-etal-11} for related works). 
Under the slow-roll approximation the nonlinear parameter 
$f_{{\rm NL}}^{{\rm equil}}$ in the equilateral configuration is
\begin{eqnarray}
f_{{\rm NL}}^{{\rm equil}} & \simeq & \frac{85}{324}\left(1-\frac{1}{c_{s}^{2}}\right)-\frac{10}{81}\,\frac{\Lambda}{\Sigma}+\frac{55}{36}\,\frac{\epsilon_{s}}{c_{s}^{2}}+\frac{5}{12}\,\frac{\eta_{s}}{c_{s}^{2}}-\frac{85}{54}\,\frac{s}{c_{s}^{2}}\nonumber \\
 &  & +\frac{5}{162}\,\delta_{F}\left(1-\frac{1}{c_{s}^{2}}\right)-\frac{10}{81}\,\delta_{\xi}\left(2-\frac{29}{c_{s}^{2}}\right)+\delta_{GX}\left[\frac{20\,(1+\lambda_{GX})}{81\epsilon_{s}}+\frac{65}{162c_{s}^{2}\epsilon_{s}}\right]\,,\label{fnleq}
\end{eqnarray}
where 
\begin{eqnarray}
\hspace{-0.5cm} &  & \Lambda \equiv F^{2}\left[\dot{\phi}H(XG_{,X}+5X^{2}G_{,XX}+2X^{3}G_{,XXX})-2(2X^{2}G_{,\phi X}+X^{3}G_{,\phi XX})/3\right],\\
\hspace{-0.5cm} &  & \Sigma\equiv\frac{w_{1}(4w_{1}w_{3}+9w_{2}^{2})}{12\Mpl^{4}}\simeq\Mpl^{2}F^{3}H^{2}\left(\delta_{X}+6\delta_{GX}-2\delta_{G\phi}+6\delta_{GX}\lambda_{GX}-2\delta_{G\phi}\lambda_{G\phi}\right)\,.
\end{eqnarray}

In the absence of the Galileon term ($\delta_{GX}=0=\delta_{G\phi}$)
one has $c_{s}^{2}\simeq1$ and $\epsilon_{s}\simeq\delta_{X}$ from 
Eqs.~(\ref{cs}) and (\ref{eps}) at linear order in slow-roll.
In this case, the expansion of $c_{s}^{2}$ up to second order gives
\begin{equation}
c_{s}^{2}\simeq1-\frac{2\delta_{\xi}(\delta_{F}-8\delta_{\xi})
(3\delta_{F}-24\delta_{\xi}-4\delta_{X})}{\delta_{X}}\,,
\end{equation}
which shows that the GB contribution can only lead to small
changes to the value $c_{s}^{2}=1$. Then the nonlinear parameter
in Eq.~(\ref{fnleq}) is approximately given by
\begin{equation}
f_{{\rm NL}}^{{\rm equil}}\simeq\frac{55}{36}\epsilon_{s}
+\frac{5}{12}\eta_{s}+\frac{10}{3}\delta_{\xi}\,,
\end{equation}
 which means that the non-Gaussianity is small for the theories with
$G=0$. However, the presence of the Galileon term can potentially
give rise to large non-Gaussianities.

%%%%%%%%%%%%%%%%%%%%%%%%%%%%%%%%%
\section{The action in the Einstein frame}
\label{Einsec} 
%%%%%%%%%%%%%%%%%%%%%%%%%%%%%%%%%

We start by considering non-minimally coupled theories in the absence
of the GB and Galileon terms ($\xi=0=G$), i.e. with actions of the form 
\begin{equation}
S=\int d^{4}x\sqrt{-g}\left[\frac{M_{{\rm pl}}^{2}}{2}F(\phi)R
+\omega(\phi)X-V(\phi)\right]\,.\label{actionsta}
\end{equation}
Since $c_{s}^{2}=1$ and $s=0$ in these theories, it follows that
\begin{eqnarray}
n_{{\rm s}}-1 & = & -2\epsilon-\delta_{Q}=
-2\epsilon-\delta_{F}-\eta_{s}\simeq-2\epsilon_{s}-\eta_{s}\,,\label{nRs}\\
n_{{\rm t}} & = & -2\epsilon-\delta_{F}\simeq-2\epsilon_{s}\,,\label{nTs}\\
r & = & 16\frac{Q}{\Mpl^{2}F}=16\epsilon_{s}\simeq-8n_{{\rm t}}\,,\label{rs}
\end{eqnarray}
where 
\begin{equation}
Q=\frac{F(2F\omega\dot{\phi}^{2}+3\Mpl^{2}\dot{F}^{2})}{(2HF+\dot{F})^{2}}\,.
\end{equation}
In the last approximate equalities of Eqs.~(\ref{nRs}), (\ref{nTs}),
and (\ref{rs}) we have used the relation $\epsilon_{s}\simeq\epsilon+\delta_{F}/2$
valid at linear order in slow roll. This follows from Eqs.~(\ref{epap})
and (\ref{eps}), i.e. $\epsilon\simeq\delta_{X}-\delta_{F}/2$ and
$\epsilon_{s}\simeq\delta_{X}$, respectively. 

It is convenient to transform the action (\ref{actionsta}), expressed in
the so called Jordan frame, into the one having a scalar field minimally coupled to
gravity (the Einstein frame), via the conformal transformation
\begin{equation}
\hat{g}_{\mu\nu}=F(\phi)g_{\mu\nu}\,.\label{ctrans}
\end{equation}
The transformed action is given by \cite{Maeda}
\begin{equation}
S_{E}=\int d^{4}x\sqrt{-\hat{g}}\left[\frac{1}{2}\Mpl^{2}\hat{R}
-\frac{1}{2}\hat{g}^{\mu\nu}\partial_{\mu}\chi\partial_{\nu}\chi-U(\chi)\right]\,,\label{Eaction}
\end{equation}
where a hat represents the quantities in the Einstein frame, and
\begin{equation}
U=\frac{V}{F^{2}}\,,\qquad\chi\equiv\int B(\phi)\, d\phi\,,\qquad 
B(\phi)\equiv\sqrt{\frac{3}{2}\left(\frac{\Mpl F_{,\phi}}{F}\right)^{2}
+\frac{\omega}{F}}\,.
\end{equation}

The following relations hold between the variables in the two frames:
\begin{equation}
d\hat{t}=\sqrt{F}\, dt\,,\qquad\hat{a}=\sqrt{F}\, a\,,\qquad
\hat{H}=\frac{1}{\sqrt{F}}\left(H+\frac{\dot{F}}{2F}\right)\,.\label{Hre}
\end{equation}
Defining the variables 
\begin{equation}
\hat{\epsilon}\equiv-\frac{1}{\hat{H}^{2}}\frac{d\hat{H}}{d\hat{t}}\,,
\qquad\hat{Q}\equiv\frac{1}{2\hat{H}^{2}}\left(\frac{d\chi}{d\hat{t}}
\right)^{2}\,,\qquad\hat{\delta}_{\hat{Q}}\equiv\frac{1}{\hat{H}
\hat{Q}}\frac{d\hat{Q}}{d\hat{t}}\,,
\end{equation}
we obtain \cite{Komatsuper,Gumjudpai}
\begin{equation}
\hat{\epsilon}=\frac{\epsilon+\delta_{F}/2}{1+\delta_{F}/2}
-\frac{\dot{\delta}_{F}}{2H(1+\delta_{F}/2)^{2}}\,,\qquad
\hat{Q}=\frac{Q}{F}\,,\qquad\hat{\delta}_{\hat{Q}}
=\frac{\delta_{Q}-\delta_{F}}{1+\delta_{F}/2}\,.
\end{equation}
Since $\hat{\epsilon}\simeq\epsilon+\delta_{F}/2$ and 
$\hat{\delta}_{\hat{Q}}\simeq\delta_{Q}-\delta_{F}$
at linear order in slow-roll, we find that Eqs.~(\ref{nRs}), (\ref{nTs}),
and (\ref{rs}) reduce to 
\begin{eqnarray}
n_{{\rm s}}-1 & \simeq & -2\hat{\epsilon}-\hat{\delta}_{\hat{Q}}\,,\label{nRs2}\\
n_{{\rm t}} & \simeq & -2\hat{\epsilon}\,,\label{nTs2}\\
r & \simeq & 16\frac{\hat{Q}}{\Mpl^{2}}=16\hat{\epsilon}\,.\label{rs2}
\end{eqnarray}
In the last equality of Eq.~(\ref{rs2}) we have used the relation
$\hat{\epsilon}=\hat{Q}/\Mpl^{2}$, which follows from the background
equation $d\hat{H}/d\hat{t}=-(d\chi/d\hat{t})^{2}/(2\Mpl^{2})$. The
results (\ref{nRs2})-(\ref{rs2}) coincide with those derived in
the Einstein frame \cite{Komatsuper,Gumjudpai}.
This equivalence is a consequence of the fact that both
the scalar and tensor spectra are unchanged under the conformal transformation
($\hat{{\cal P}}_{{\rm s}}={\cal P}_{{\rm s}}$ 
and $\hat{{\cal P}}_{{\rm t}}={\cal P}_{{\rm t}}$) \cite{nonminimalper}.

Under the slow-roll conditions ($|d^{2}\chi/d\hat{t}^{2}|\ll|3\hat{H}d\chi/d\hat{t}|$
and $(d\chi/d\hat{t})^{2}/2\ll U$) the background equations are 
approximately given by
\begin{equation}
3\Mpl^{2}\hat{H}^{2}\simeq U\,,\qquad
3\hat{H}\frac{d\chi}{d\hat{t}}\simeq-U_{,\chi}\,.
\label{Einback}
\end{equation}
We then have 
\begin{equation}
\hat{\epsilon}=\frac{\hat{Q}}{\Mpl^{2}}\simeq\frac{\Mpl^{2}}{2}
\left(\frac{U_{,\chi}}{U}\right)^{2}\,,\qquad\hat{\delta}_{\hat{Q}}\simeq
2\Mpl^{2}\left[\left(\frac{U_{,\chi}}{U}\right)^{2}-\frac{U_{,\chi\chi}}{U}\right]\,.
\end{equation}
The observables (\ref{nRs2})-(\ref{rs2}) can be explicitly written as 
\begin{eqnarray}
n_{{\rm s}}-1 & \simeq & -3\Mpl^{2}\left(\frac{U_{,\chi}}{U}\right)^{2}+2\Mpl^{2}\frac{U_{,\chi\chi}}{U}\simeq\frac{\Mpl^{2}}{B^{2}}\left[2\frac{V_{,\phi\phi}}{V}-3\frac{V_{,\phi}^{2}}{V^{2}}-4\frac{F_{,\phi\phi}}{F}+4\frac{V_{,\phi}}{V}\frac{F_{,\phi}}{F}-2\frac{B_{,\phi}}{B}\left(\frac{V_{,\phi}}{V}-2\frac{F_{,\phi}}{F}\right)\right],\label{nRex}\\
r & \simeq & -8n_{{\rm t}}\simeq8\Mpl^{2}\left(\frac{U_{,\chi}}{U}\right)^{2}\simeq8\frac{\Mpl^{2}}{B^{2}}\left(\frac{V_{,\phi}}{V}-2\frac{F_{,\phi}}{F}\right)^{2}\,.\label{rex}
\end{eqnarray}
In the Jordan frame the number of e-foldings from the time $t$ (with
the field value $\phi$) to the time $t_f$ at the end of inflation
(with the field value $\phi_f$) is given by 
\begin{equation}
N=\int_{t}^{t_{f}}Hdt=\int_{\hat{t}}^{\hat{t}_{f}}
\hat{H}d\hat{t}+\frac{1}{2}\ln\frac{F}{F_{f}}\,,\label{efold0}
\end{equation}
where $F_f \equiv F(\phi_f)$. Note that in the last equality
we have used Eq.~(\ref{Hre}). The scales relevant to the
CMB temperature anisotropies correspond to $N=50$-60 \cite{Leach}. 
The number of e-foldings in the Einstein frame should be equivalent to that in the Jordan
frame by properly choosing some reference length scale \cite{Catena}.
Using the slow-roll approximation in the Einstein frame, the frame-independent
quantity (\ref{efold0}) can be written as 
\begin{equation}
N\simeq\int_{\chi_{f}}^{\chi}\frac{U}{\Mpl^{2}U_{,\chi}}d\chi
+\frac{1}{2}\ln\frac{F}{F_{f}}\,,\label{efold2}
\end{equation}
which we will use in the following sections.

%%%%%%%%%%%%%%%%%%%%%%%%%%%%%%%%%%%%%%%
\section{Inflation with nonminimal coupling and field
coupling with the kinetic term}
\label{nonminimalsec} 
%%%%%%%%%%%%%%%%%%%%%%%%%%%%%%%%%%%%%%%

In this section we study, in turn, models with the nonminimal coupling 
$\zeta \phi^2 R/2$ and the non-canonical kinetic term $\omega (\phi)X$.
These models are described by the action 
\begin{equation}
S=\int d^{4}x\sqrt{-g}\left[\frac{\Mpl^{2}}{2}R
-\frac{1}{2}\zeta\phi^{2}R+\omega(\phi)X-V(\phi)\right]\,.
\end{equation}
In this case the function $F$ is given by 
\begin{equation}
F=1-\zeta x^{2}\,,\qquad x\equiv\phi/\Mpl\,.
\end{equation}
Note that in our notation the conformal coupling corresponds 
to $\zeta=1/6$.
For the canonical field with $\omega(\phi)=1$, the observational
constraints were studied for the chaotic potential of the type (\ref{chaoticpo})
by using the WMAP 1yr data combined with the large-scale structure
data \cite{Gumjudpai}. Recently the observational compatibility of 
this type of potential as well as $V(\phi)=\lambda (\phi^{2}-v^{2})^{2}/4$
was examined in Ref.~\cite{Linde} by using the WMAP 7yr data.
The latter potential appears in the context of Higgs inflation with the electroweak scale
$v\sim10^{3}$~GeV \cite{Bez}. If the nonminimal coupling is negative
with $|\zeta|\gg1$, it is possible to use the Higgs field as an inflaton
because the self coupling $\lambda$ can be of the order of 0.01-0.1 from 
the WMAP normalization.
Since the field $\phi$ is much larger than the electroweak scale
during inflation, the observational prediction of the potential 
$V(\phi)=\lambda (\phi^{2}-v^{2})^{2}/4$
is very similar to that of the potential (\ref{chaoticpo}) with $p=4$.

In this work we shall take into account the non-canonical kinetic term 
$\omega(\phi) X$ in addition to the nonminimal coupling $\zeta \phi^2 R/2$. 
We provide general formulae for $n_{\rm s}$, $r$, 
and $n_{\rm t}$ in terms of the function of $x=\phi/\Mpl$
and then apply them to the cases where $\omega(\phi)= {\rm constant}$ 
and where the exponential coupling $\omega (\phi)=e^{\mu \phi/\Mpl}$ 
is present. In the Einstein frame this potential takes the form
\begin{equation}
U=\frac{V_{0}x^{p}}{(1-\zeta x^{2})^{2}}\,.\label{Einspoten}
\end{equation}
For $p<4$ this has a local maximum at $x=\sqrt{p/[(4-p)|\zeta|]}$
and hence the nonminimal coupling makes it more difficult to realise
inflation. If $p=4$ the potential (\ref{Einspoten}) is asymptotically
flat in the region $\phi\gg\Mpl$. If $p>4$ the potential does not
possess a local maximum, but for $p>5+\sqrt{13}$ 
inflation does not occur.

From Eqs.~(\ref{nRex}) and (\ref{rex}) it follows that 
\begin{eqnarray}
n_{{\rm s}}-1 & \simeq & -\frac{1}{[\omega+(6\zeta-\omega)\zeta x^{2}]^{2}x^{2}}\biggl\{(p-4)^{2}(6\zeta-\omega)(\zeta x^{2})^{3}+(24\omega-14p\omega+3p^{2}\omega+24p\zeta-12p^{2}\zeta)(\zeta x^{2})^{2}\nonumber \\
 &  & +(-8\omega+4p\omega-3p^{2}\omega+24p\zeta+6p^{2}\zeta)\zeta x^{2}+p\omega(p+2)-\mu\omega x(1-\zeta x^{2})^{2}[(p-4)\zeta x^{2}-p]\biggr\}\,,\label{nRex2}\\
r & \simeq & -8n_{{\rm t}}\simeq\frac{8[p+(4-p)\zeta x^{2}]^{2}}{x^{2}[\omega+(6\zeta-\omega)\zeta x^{2}]}\,,\label{rex2}
\end{eqnarray}
where 
\begin{equation}
\mu\equiv\frac{\Mpl\omega_{,\phi}}{\omega}\,.
\end{equation}
For the dilatonic coupling $\omega(\phi)=e^{\mu\phi/\Mpl}$
the parameter $\mu$ is constant. 
Using the approximate equations (\ref{Einback}), 
the scalar power spectrum is given by 
\begin{equation}
{\cal P}_{\rm s} \simeq \frac{U^3}{12\pi^2 \Mpl^6 U_{,\chi}^2}
=\frac{V_0}{12 \pi^2 \Mpl^4} \frac{x^{p+2} [6 \zeta^2 x^2+\omega
(1-\zeta x^2)]}{(1-\zeta x^2)^2 [p+(4-p)\zeta x^2]^2}\,.
\label{Psam}
\end{equation}
The WMAP normalization corresponds to ${\cal P}_{\rm s} \simeq 2.4 \times 10^{-9}$
at the scale $k=0.002$\,Mpc$^{-1}$.
In the following we shall first consider
the nonminimally coupled theories with $\mu=0$ and then proceed to
the case in which the dilatonic kinetic term 
$e^{\mu \phi/\Mpl}X$ is present.

\subsection{Effect of the nonminimal coupling $\zeta \phi^2 R/2$
with constant $\omega$}

We first discuss the effect of the nonminimal coupling for 
the theories with 
\begin{equation}
\mu=0\,,
\end{equation}
in which case $\omega$ is constant. 
Introducing a new field $\varphi=\sqrt{\omega}\phi$
the kinetic term $\omega X$ reduces to the canonical form
$-g^{\mu \nu} \partial_{\mu} \varphi \partial_{\nu} \varphi/2$.
Then the nonminimal coupling $\zeta \phi^2 R/2$ can be written 
as $\tilde{\zeta} \varphi^2 R/2$, where $\tilde{\zeta}=\zeta/\omega$.
The potential $V(\phi)=V_0(\phi/\Mpl)^p$ for the scalar field $\varphi$, 
takes the power-law form 
$V=\tilde V_0(\varphi/\Mpl)^p$, where $\tilde V_0=V_0/\omega^{p/2}$.
This means that these theories reduce to nonminimally 
coupled theories with $\omega=1$ in terms of the field $\varphi$.
The ratio $\tilde{\zeta}=\zeta/\omega$ characterizes the effect of 
the nonminimal coupling on the inflationary observables $n_{\rm s}$,
$n_{\rm t}$, and $r$, and $\tilde V_0=V_0/\omega^{p/2}$ sets 
the scale for the scalar power spectrum.

{}From Eq.~(\ref{efold2}) the number of e-foldings is given by 
\begin{eqnarray}
&  & N\simeq-\frac{1}{4\zeta}\ln\left|\frac{(p-4)\zeta x_{f}^{2}-p}
{(p-4)\zeta x^{2}-p}\right|^{\frac{3p\zeta-2\omega}{p-4}}
-\frac{1}{4}\ln\left|\frac{1-\zeta x^{2}}{1-\zeta x_{f}^{2}}\right|
\qquad(p\neq4)\,,\label{Nap0}\\
&  & N\simeq\frac{\omega-6\zeta}{8}(x^{2}-x_{f}^{2})
-\frac{1}{4}\ln\left|\frac{1-\zeta x^{2}}{1-\zeta x_{f}^{2}}\right|
\qquad(p=4)\,,\label{Nap}
\end{eqnarray}
where $x_f \equiv \phi_f/\Mpl$.
The result (\ref{Nap}) can be also reproduced by taking the limit $p\to4$
in Eq.~(\ref{Nap0}). 
We identify the end of inflation by the condition
$\hat{\epsilon}=1$, which gives  
\begin{equation}
x_{f}^{2}=\frac{\omega-\zeta p(4-p)-\sqrt{(\omega-2p\zeta)
(\omega-6p\zeta)}}{\zeta[\zeta(4-p)^{2}+2(\omega-6\zeta)]}\,.
\label{psif}
\end{equation}

Let us consider the limits where $|\zeta/\omega| \ll 1$. We implicitly assume
that $\omega$ is not different from the order of 1. We expand the
right hand side of Eqs.~(\ref{Nap0}) and (\ref{Nap}) up to first
order in $\zeta$ and then solve them for $x$ by 
using Eq.~(\ref{psif}). This gives 
\begin{equation}
x^{2} \simeq \frac{p(p+4N)}{2\omega} \left[ 1
-\frac{8(p-4)N^2+4p (p-6)N+p^2(p-8)}{2(p+4N)}
\frac{\zeta}{\omega} \right]\,,
\label{psi1}
\end{equation}
which is valid for both $p \neq 4$ and $p=4$.
The spectral index (\ref{nRex2}) and the tensor-to-scalar ratio
(\ref{rex2}) are approximately given by 
\begin{eqnarray}
n_{{\rm s}}-1 & \simeq & -{\frac{2(p+2)}{p+4N}}
\left[1-{\frac{4(p-2)(p-12){N}^{2}+2p({p}^{2}-12\,{p}+28)N+p^{2}(12-p)}
{\left(p+4\, N\right)\left(p+2\right)}} \frac{\zeta}{\omega}\right]\,,\label{nRap}\\
r & \simeq & {\frac{16p}{p+4\, N}}\left[1-{\frac{2N\bigl(2(p-12)N+p(p-10)\bigr)}
{p+4N}} \frac{\zeta}{\omega}\right]\,.
\label{rap}
\end{eqnarray}
This shows that the effect of the nonminimal coupling appears 
in terms of the ratio $\zeta/\omega$.

Substituting Eq.~(\ref{psi1}) into Eq.~(\ref{Psam}) and expanding it up to 
first order in $\zeta$, it follows that 
\begin{eqnarray}
{\cal P}_{\rm s} &\simeq& \frac{{\tilde V}_0}{\Mpl^4} \frac{p+4N}{24 \pi^2 p}
\left[ \frac{p(p+4N)}{2} \right]^{p/2}
\left[ 1-\frac{p^4+(4N-12)p^3+(8N^2-64N)p^2+(80N-112N^2)p+192N^2}
{4(p+4N)} \frac{\zeta}{\omega} \right] \nonumber \\
&=& 2.4 \times 10^{-9}\,,
\label{Pscon}
\end{eqnarray}
around $N=55$. 
In the absence of the nonminimal coupling ($\zeta=0$) 
the WMAP normalization for the canonical scalar field $\varphi$
gives $m \simeq 6.8 \times 10^{-6}\Mpl$
for $p=2$ (where $\tilde V_0=m^2 \Mpl^2/2$) and 
$\lambda \simeq 2.0 \times 10^{-13} $ for 
$p=4$ (where $\tilde V_0=\lambda \Mpl^4/4$).
If $\zeta \neq 0$, then the inside of the last parenthesis
in Eq.~(\ref{Pscon}) is approximately 
given by $1+4\zeta/\omega$ for $p=2$ and 
$1+460 \zeta/\omega$ for $p=4$.
As long as $|\zeta/\omega| \ll 1$, the order of ${\tilde V}_0$ is not 
subject to change by the presence of the nonminimal coupling.

In the following we derive the numerical values of $n_{\rm s}$
and $r$ for $p=2$ and $p=4$ separately
to compare the models with observations.

\subsubsection{$p=2$}

In order to obtain the theoretical values of $n_{{\rm s}}$ and $r$ 
for $p=2$, we numerically solve the background equations of motion
in the Jordan frame by identifying the end of inflation 
under the condition (\ref{psif}). 
We derive the numerical values of $x$ corresponding
to the number of e-foldings $N=55$ and then evaluate $n_{{\rm s}}$
and $r$ by using the formulas (\ref{nRex2}) and (\ref{rex2}).

\begin{figure}
\begin{centering}
\includegraphics[width=5.5in,height=4in]{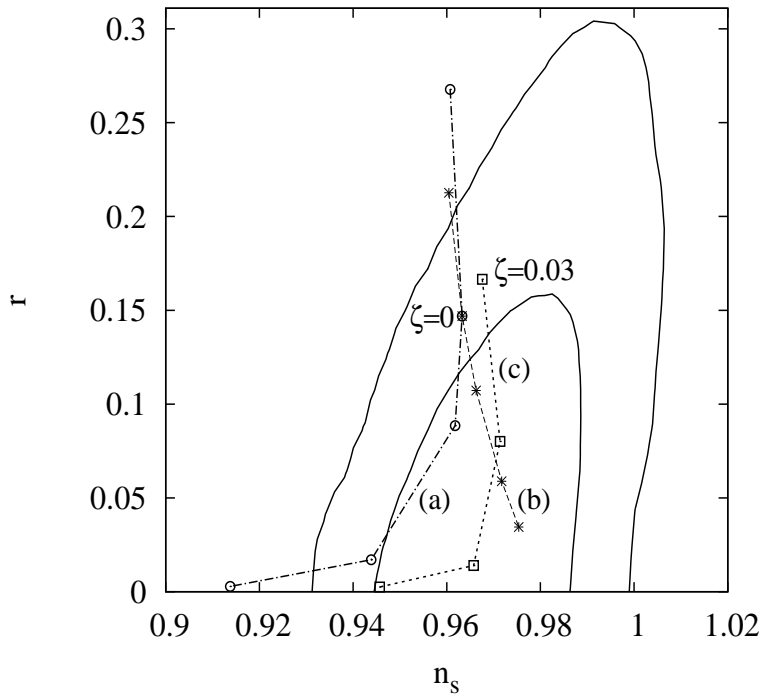} 
\par\end{centering}
\caption{1$\sigma$ and 2$\sigma$ observational contours
in the $(n_{{\rm s}},r)$
plane constrained by the joint data analysis of WMAP 7yr, BAO, and
HST with the pivot scale $k_{0}=0.002$\,Mpc$^{-1}$.
Shown also are the theoretical predictions 
for the potential $V(\phi)=m^2 \phi^2/2$
for $N=55$ in three cases: 
(a) constant $\omega$ (i.e.~$\mu=0$) 
in the presence of the nonminimal coupling $\zeta \phi^2 R/2$ with 
$\zeta/\omega=0.001, 0,-0.001,-0.005,-0.01$ (from top to bottom), 
(b) the exponential coupling $e^{\mu\phi/\Mpl}X$
with $\mu=-0.05,0, 0.1,1,10$ (from top to bottom)
in the absence of the nonminimal coupling, and
(c) the exponential coupling $e^{\phi/\Mpl}X$ (i.e. $\mu=1$)
in the presence of the nonminimal coupling with 
$\zeta=0.03, 0.01, -0.05,-0.1$ (from top to bottom).}
\centering{}\label{fig1} 
\end{figure}

In Fig.~\ref{fig1} we show the $1\sigma$ and $2\sigma$ observational
contours constrained by the joint data analysis of WMAP 7yr \cite{WMAP7}, 
Baryon Acoustic Oscillations (BAO) \cite{BAO}, 
and the Hubble constant measurement (HST) \cite{HST}. 
This is derived by varying the two parameters $n_{\rm s}$
and $r$ with the consistency relation $r=-8n_{\rm t}$
[see Eq.~(\ref{rs})].
Since the runnings of scalar and tensor spectral indices
are suppressed to be of the order of $\epsilon^2$, 
they are set to be 0 in the likelihood analysis. 
These results are valid for the theories with $\xi=0=G$.

In the limit $|\zeta|\ll1$, Eqs.~(\ref{nRap}) and (\ref{rap}) give
\begin{eqnarray}
n_{{\rm s}}-1 &\simeq&-\frac{4}{2N+1} \left[1
-\frac{4N+5}{2N+1} \frac{\zeta}{\omega}
+\frac { 2\left(104 {N}^{4}+160 {N}^{3}+84 {N}^{2}-30\,N-9 \right)}
{3 \left( 2N+1 \right)^{2}}\frac{\zeta^2}{\omega^2} \right]\,,
\label{nsp=2}\\
r & \simeq & \frac{16}{2N+1}\left[1+{\frac {4N 
\left( 5N+4 \right)}{2N+1}}\,\frac\zeta\omega\right].
\label{rp=2}
\end{eqnarray}
For the scalar index we have included the second-order correction 
in $\zeta/\omega$ because the dominant contribution to the 
first-order term in $\zeta/\omega$ in Eq.~(\ref{nRap}) 
vanishes for $p=2$.
In the absence of the nonminimal coupling
($\zeta=0$) one has $n_{{\rm s}}=0.964$ and $r=0.144$ 
for $N=55$, which is inside the $2\sigma$ observational bound
(see Fig.~\ref{fig1}).
A positive nonminimal coupling leads to 
an increase of $r$ relative to the case $\zeta=0$.
Since $r$ is bounded from above observationally, this puts
an upper bound on the positive value of $\zeta$.
The negative nonminimal coupling gives rise to the deviation
from the scale-invariant spectrum ($n_{{\rm s}}=1$) and 
the decrease of $r$. 

From the observational constraints on $n_{\rm s}$ we can place 
the bound on the negative nonminimal coupling.
We find that the ratio $\zeta/\omega$ is constrained to be 
\begin{equation}
-7.0\times10^{-3}<\zeta/\omega<7.0 \times 10^{-4}
\qquad (95\,\%~{\rm CL})\,,
\label{p=2con}
\end{equation}
which agrees with that derived in Ref.~\cite{Linde} for $\omega=1$.
The lower bound in Eq.~(\ref{p=2con}) is slightly tighter than the constraint 
$\zeta>-1.1\times10^{-2}$ (with $\omega=1$) \cite{Gumjudpai} obtained 
by using the WMAP 1yr data combined with
the large-scale structure data.

%%%%%%%%%%%%%%%%%%%%%%%%%%%%%
%%%%%%%%%%%%%%%%%%%%%%%%%%%%

\subsubsection{$p=4$}

%%%%%%%%%%%%%%%%%%%%%%%%%%%%%
\begin{figure}
\begin{centering}
\includegraphics[width=5.5in,height=4in]{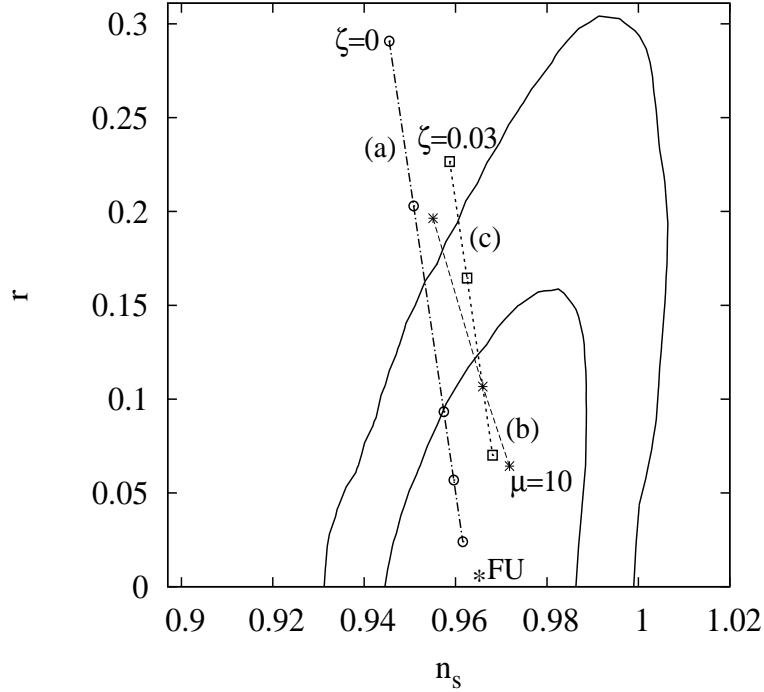} 
\par\end{centering}
\caption{The same observational constraints as shown in Fig.~\ref{fig1},
with the theoretical prediction of the potential $V(\phi)=\lambda \phi^4/4$
for $N=55$. Each curve corresponds to 
(a) constant $\omega$ (i.e.~$\mu=0$) 
in the presence of the nonminimal coupling $\zeta \phi^2 R/2$ with 
$\zeta/\omega=0,-0.001,-0.005,-0.01,-0.03$ (from top to bottom), 
(b) the exponential coupling $e^{\mu\phi/\Mpl}X$
with $\mu=0.1,1,10$  (from top to bottom)
in the absence of the nonminimal coupling, and
(c) the exponential coupling $e^{\phi/\Mpl}X$ (i.e. $\mu=1$)
in the presence of the nonminimal coupling with 
$\zeta =0.03, 0.02, -0.03$ (from top to bottom).
The label ``FU'' corresponds to the Fakir-Unruh 
scenario \cite{Fakir} with $\zeta \to -\infty$.}
\centering{}\label{fig2} 
\end{figure}
%%%%%%%%%%%%%%%%%%%%%%%%%%%%

We proceed to the case of the self-coupling inflaton potential  
$V(\phi)=\lambda \phi^4/4$. 
In the regime $|\zeta/\omega|\ll1$ Eqs.~(\ref{nRap})
and (\ref{rap}) give 
\begin{eqnarray}
n_{{\rm s}}-1 & \simeq & -\frac{3}{N+1}\left[1+{\frac { 4\left( 2{N}^{2}+N-4
 \right)}{3(N+1)}}\,\frac\zeta\omega\right],\\
r & \simeq & \frac{16}{N+1}\left[1+{\frac {4N \left( 2N+3 \right) 
}{N+1}}\,\frac\zeta\omega\right]\,.
\end{eqnarray}
In the absence of the nonminimal coupling one has $n_{{\rm s}}=0.946$
and $r=0.286$ for $N=55$, which is outside the $2\sigma$ observational
bound (see Fig.~\ref{fig2}). The presence of the negative nonminimal
coupling leads to the increase of $n_{\rm s}$, whereas $r$ gets smaller.
Hence it is possible for the self-coupling inflaton potential to be 
consistent with observations.
{}From the joint data analysis of WMAP 7yr, BAO, and HST the 
nonminimal coupling is constrained to be 
\begin{equation}
\zeta/\omega<-2.0 \times 10^{-3} \qquad 
(95\,\%~{\rm CL})\,,
\end{equation}
which is tighter than the bound $\zeta<-3.0 \times 10^{-4}$ 
(with $\omega=1$) derived in Ref.~\cite{Gumjudpai}.

In another limit where $|\zeta/\omega| \to \infty$, inflation
is realised by the flat potential $U$ in the Einstein frame
in the regime $x \gg 1$.
In this case one has $x_{f}^2\simeq-2\sqrt{3}/(3\zeta)$ 
and $N\simeq-3\zeta x^{2}/4$
from Eqs.~(\ref{psif}) and (\ref{Nap}), respectively. From Eqs.~(\ref{nRex2})
and (\ref{rex2}) the leading contributions to $n_{{\rm s}}$ and
$r$ in the regime $N\gg1$ are 
\begin{eqnarray}
n_{{\rm s}}-1 & \simeq & -2/N\,,\label{nslimit}\\
r & \simeq & 12/N^{2}\,.\label{rlimit}
\end{eqnarray}
As long as $|\zeta|$ is sufficiently large relative to $\omega$,
the effect of the term $\omega$ appears only as the next order corrections
to (\ref{nslimit}) and (\ref{rlimit})
with the order of $\omega/(\zeta N^{2})$. For $N=55$ one has $n_{{\rm s}}=0.964$
and $r=0.004$ from (\ref{nslimit}) and (\ref{rlimit}),
which are well inside the $1\sigma$ observational bound.

In the regime $|\zeta/\omega| \gg 1$ the power spectrum (\ref{Psam}) 
reduces to ${\cal P}_{\rm s} \simeq \lambda N^2/(72 \pi^2 \zeta^2)$, 
so that the WMAP normalization 
${\cal P}_{\rm s} \simeq 2.4 \times 10^{-9}$ at $N=55$ gives
\begin{equation}
\lambda/\zeta^2 \simeq 5.6 \times 10^{-10}\,.
\end{equation}
For large negative nonminimal couplings, such as $\zeta \sim -10^{4}$, 
the self coupling $\lambda$ can be of the order of $10^{-2}$.
This property was used in the context of Higgs inflation.

\subsection{Effect of the non-canonical kinetic term $e^{\mu \phi/\Mpl}X$ with $\zeta=0$}

Let us consider the case in which the field-dependent coupling 
$\omega(\phi)=e^{\mu\phi/M_{{\rm pl}}}$ with the kinetic energy $X$
is present, without taking into account the nonminimal coupling ($\zeta=0$).
After the field settles down to the potential minimum, $\phi=0$, 
the coupling $\omega(\phi) \to 1$ and one recovers
the standard kinetic energy $X$.

The number of e-foldings (\ref{efold2}) is given by 
\begin{equation}
N=\frac{1}{p\mu^{2}}\left[(\mu x-1)e^{\mu x}-(\mu x_{f}-1)e^{\mu x_{f}}\right]\,.
\label{Nderi}
\end{equation}
Since $\hat{\epsilon}=p^{2}/(2x^{2}\omega)$, we can
estimate $x_{f}$ by setting $\hat{\epsilon}=1$: 
\begin{equation}
x_{f}^{2}\, e^{\mu x_{f}}=p^{2}/2\,,\qquad{\rm or}
\qquad x_{f}=\frac{2W(\sqrt{2}|\mu|p/4)}{\mu},\label{xf}
\end{equation}
where $W$ is the Lambert's $W$ function \cite{Lambert}.
The scalar spectral index
(\ref{nRex2}) and the tensor-to-scalar ratio (\ref{rex2}) are 
\begin{eqnarray}
n_{{\rm s}}-1 & = & -\frac{p}{x^{2}e^{\mu x}}\left(p+2+\mu x\right)\,,\label{nski}\\
r & = & \frac{8p^{2}}{x^{2}e^{\mu x}}\,.
\label{rki}
\end{eqnarray}

In the limit $|\mu|\ll1$, using Eq.~(\ref{xf}), one can rewrite
Eq.~(\ref{Nderi}) in the form 
\begin{equation}
N\simeq{\frac{2x^{2}-{p}^{2}}{4p}}+\frac{\left(8x^{3}
+\sqrt{2}p^{3}\right)\mu}{24p}\,,
\end{equation}
which can be solved for $x$, as 
\begin{equation}
x^{2}\simeq\frac{p^{2}}{2}+2pN-\frac{\mu}{12}
\left[\sqrt{2}\, p^{3}+(2p^{2}+8pN)^{3/2}\right]\,.
\end{equation}
By replacing this relation into Eqs.~(\ref{nski}) and (\ref{rki}), we obtain
\begin{eqnarray}
n_{{\rm s}}-1 & \simeq & -\frac{p+2}{2N}
\left[1-\frac{(p-1)\mu\sqrt{2pN}}{3(p+2)}\right]\,,\label{nski2}\\
r & \simeq & \frac{4p}{N}\left(1-\frac{\mu\sqrt{2pN}}{3}\right)\,,
\end{eqnarray}
which are valid up to the first order in $\mu$. The presence of
the positive $\mu$ leads to the approach to the scale-invariant
spectrum, whereas $r$ gets smaller. In Figs.~\ref{fig1} and \ref{fig2}
we plot the theoretical values of $n_{{\rm s}}$ and $r$ 
in the ($n_{{\rm s}},r$) plane for $p=2$ and $p=4$, 
respectively, with several different
values of $\mu$. 
These are derived numerically by integrating the
background equations without using the approximation given above
(because the approximation loses its validity for $\mu \gtrsim 1$).
Interestingly the models with large positive values of $\mu$ can
be favoured observationally. On the other hand, the models with negative
$\mu$ lead to the deviation from the observationally allowed region.
The joint observational constraints
from WMAP 7yr, BAO, and HST give the following bounds on $\mu$: 
\begin{eqnarray}
 &  & \mu>-0.04\quad(95\%\,{\rm CL})\quad{\rm for}\quad p=2\,,\\
 &  & \mu>0.2\quad(95\%\,{\rm CL})\quad{\rm for}\quad p=4\,.
\end{eqnarray}

Let us now consider the limit where $\mu\gg1$. In this regime the condition
$\mu x\gg1$ is satisfied, so that $N\simeq xe^{\mu x}/(p\mu)$ from
Eq.~(\ref{Nderi}). Then the scalar index (\ref{nski}) and the tensor-to-scalar
ratio (\ref{rki}) reduce to 
\begin{eqnarray}
n_{{\rm s}}-1 & \simeq & -\frac{1}{N}\,,\label{nski3}\\
r & \simeq & \frac{8p}{N}\frac{1}{\mu x}\,.
\label{rki3}
\end{eqnarray}
For a given $N$, $\mu x$ increases for larger $\mu$.
This means that, in the limit $\mu\gg1$, $n_{{\rm s}}$ and $r$
approach $n_{{\rm s}}\to1-1/N \simeq 0.982$ (for $N=55$) 
and $r\to0$, respectively, which is inside the $1\sigma$ observational 
bound. We have also confirmed numerically that inflation is followed
by a reheating phase with oscillations of $\phi$.

\subsection{Combined effects of the nonminimal coupling $\zeta \phi^2 R/2$ and  
the non-canonical kinetic term $e^{\mu \phi/\Mpl}X$}

Finally we consider the case in which both the nonminimal 
coupling $\zeta \phi^2 R/2$ and the non-canonical kinetic term $e^{\mu \phi/\Mpl}X$
are taken into account. 
Since it is difficult to derive an analytic form for the 
number of e-foldings $N$, we solve the background equations numerically 
to identify the values $x$ corresponding to $N=55$ 
before $x=x_f$ which happens when $\hat{\epsilon}=1$.
We then use the formulae (\ref{nRex}) and (\ref{rex}) to evaluate $n_{\rm s}$
and $r$ for given values of $p$, $\mu$, and $\zeta$.

In Fig.~\ref{fig1} we plot the numerical values of $n_{\rm s}$ and $r$
in the two-dimensional plane for $p=2$ and $\mu=1$ with 
$\zeta=0.03, 0.01, -0.05, -0.1$.
The presence of the term $e^{\mu \phi/\Mpl}X$ with $\mu>0$ 
leads to the compatibility of the nonminimally coupled models 
with larger values of $|\zeta|$ than those for $\mu=0$ and 
$\omega=1$.
When $\mu=1$ we find that the nonminimal coupling is 
constrained to be 
\begin{equation}
 -0.12< \zeta<0.035
 \quad(95\%\,{\rm CL})\,,
 \label{p2com}
\end{equation}
which is wider than the range (\ref{p=2con}).
For values of $|\zeta|$ larger than the bounds given by (\ref{p2com})
the effect of the nonminimal coupling is more important 
than that of the non-canonical kinetic term.

For $p=4$ and $\mu=0$ a positive nonminimal coupling
is not allowed observationally because both $|n_{\rm s}-1|$
and $r$ tend to be larger than those for $\zeta=0$.
However, the non-canonical kinetic term with $\mu>0$ 
allows the compatibility of the positive nonminimally 
coupled models with observations (see Fig.~\ref{fig2}).
If $\mu=1$, $\zeta$ is constrained to be 
\begin{equation}
\zeta<0.025
\quad(95\%\,{\rm CL})\,.
\label{p4com}
\end{equation}
For $\mu=1$ the models with $\zeta<0$
are within the $1\sigma$ observational bound.
In the limit of the largely negative nonminimal coupling ($|\zeta| \gg 1$), 
the scalar index and the tensor-to-scalar ratio are given 
by Eqs.~(\ref{nslimit}) and (\ref{rlimit}).

For $\mu$ larger than the order of 1 the effect of the 
non-canonical kinetic term tends to be more important. 
In the limit that $\mu \gg 1$ with a finite value of $\zeta$ (where $|\zeta| \lesssim 1$), 
$n_{\rm s}$ and $r$ approach the values given 
by Eqs.~(\ref{nski3}) and (\ref{rki3}).

%%%%%%%%%%%%%%%%%%%
\section{Inflation in the context of Brans-Dicke theories}
\label{BDsec} 
%%%%%%%%%%%%%%%%%%%

Let us proceed to Brans-Dicke (BD) theory \cite{Brans} with the action
\begin{equation}
S=\int d^{4}x\sqrt{-g}\left[\frac{1}{2}\Mpl\phi R
+\frac{\Mpl}{\phi}\omega_{{\rm BD}}X-V(\phi)\right]\,,
\label{Jframe}
\end{equation}
where $\omega_{{\rm BD}}$ is the BD parameter. 
Here we have introduced the reduced Planck mass $\Mpl$
in the first two terms, so that the field $\phi$ has 
a dimension of mass.
Under the conformal transformation (\ref{ctrans}) 
we obtain the action (\ref{Eaction})
in the Einstein frame with
\begin{equation}
F=\frac{\phi}{\Mpl}=e^{\mu\chi/\Mpl}\,,\qquad 
U=e^{-2\mu\chi/\Mpl}\, V\,,
\end{equation}
where 
\begin{equation}
\mu\equiv1/\sqrt{3/2+\omega_{{\rm BD}}}\,.\label{mudef}
\end{equation}
The integration constant for the field $\chi$ is chosen such that
$\chi=0$ corresponds to $\phi=\Mpl$.

\subsection{Case of the power-law potential}

Let us consider the power-law potential (\ref{chaoticpo}) in the Jordan frame.
In the Einstein frame the potential is given by 
\begin{equation}
U(\chi)=V_{0}e^{\lambda\chi/\Mpl}\,,\qquad
\lambda \equiv \frac{p-2}{\sqrt{3/2+\omega_{{\rm BD}}}}\,.\label{exponential}
\end{equation}
For the BD parameter of the order of 1 inflation does not occur unless
$p$ is close to 2. However, for $\omega_{{\rm BD}} \gg 1$, 
it is possible to realise $|\lambda| \ll 1$ even if $p$ is away from $2$.

On using Eqs.~(\ref{nRex}) and (\ref{rex}) for the potential 
(\ref{exponential}), it follows that 
\begin{eqnarray}
n_{{\rm s}}-1 & = & -\lambda^{2}\,,\\
r & = & -8n_{\rm t}=8\lambda^{2}\,.
\end{eqnarray}
The CMB likelihood analysis using the data of WMAP 7yr \cite{WMAP7} 
combined with BAO \cite{BAO} and HST \cite{HST}
gives the following bound on $\lambda$ \cite{Ohashi}:
\begin{equation}
0.09<\lambda<0.23\quad(95\%~{\rm CL}).
\end{equation}
This translates into the constraint on $\omega_{{\rm BD}}$: 
\begin{equation}
19(p-2)^{2}-3/2<\omega_{{\rm BD}}<123(p-2)^{2}-3/2\,.\label{omecon}
\end{equation}
The reason why the $p=2$ case (i.e. $\lambda=0)$ is disfavoured
is that the Harrison-Zel'dovich spectrum ($n_{\rm s}=1$ and $r=0$)
is in tension with observations \cite{WMAP7}.
If $p=4$, Eq.~(\ref{omecon}) gives the bound $75<\omega_{{\rm BD}}<491$.

The exponential potential in the Einstein frame does not lead to the
end of inflation, so the above scenario has to be modified in a way
that the potential has a minimum to lead to a successful reheating.
In the following we shall consider the modification of the power-law
potential in the Jordan frame, such that inflation ends as in the
Starobinsky's $f(R)$ model \cite{Star80}.

\subsection{Models including the Starobinsky's $f(R)$ scenario}

The $f(R)$ theory with the action 
\begin{equation}
S=\int d^{4}x\sqrt{-g}\frac{\Mpl^{2}}{2}f(R)\,,\label{actionfR}
\end{equation}
is equivalent to BD theory with $\omega_{{\rm BD}}=0$ \cite{Ohanlon}.
In fact the action (\ref{actionfR}) can be written as 
\begin{equation}
S=\int d^{4}x\sqrt{-g}\left[\frac{\Mpl^{2}}{2}F(\phi)R-V(\phi)\right]\,,
\label{actionfR2}
\end{equation}
where 
\begin{equation}
F=\frac{\phi}{\Mpl}=\frac{\partial f}{\partial R}\,,\qquad V(\phi)=\frac{\Mpl^{2}}{2}\left(R\frac{\partial f}{\partial R}-f\right)\,.
\end{equation}

In Starobinsky's model $f(R)=R+R^{2}/(6M^{2})$, 
we have $R=3M^{2}(\phi/\Mpl-1)$ and 
\begin{equation}
V(\phi)=\frac{3M^{2}}{4}(\phi-\Mpl)^{2}\,.\label{Stapo}
\end{equation}
We consider the following more general potential 
\begin{equation}
V(\phi)=V_{0}(\phi-\Mpl)^{p}\,,
\end{equation}
with arbitrary values of $\omega_{{\rm BD}}$, so that the model
$f(R)=R+R^{2}/(6M^{2})$ is covered as a special case with $p=2$ and
$\omega_{{\rm BD}}=0$. The potential in the Einstein frame reads
\begin{equation}
U=V_{0}{\Mpl}^{p}e^{(p-2)\mu\chi/\Mpl}\left(1-e^{-\mu\chi/\Mpl}\right)^{p}\,,
\label{poEin}
\end{equation}
where $\mu$ is defined in Eq.~(\ref{mudef}). 
For $|\omega_{{\rm BD}}|\sim{\cal O}(1)$,
i.e. $\mu\sim{\cal O}(1)$, inflation occurs in the regime $\chi\gg\Mpl$
(including the Starobinsky's $f(R)$ model). The behaviour of the potential
(\ref{poEin}) depends on the values of $p$: 
\begin{itemize}
\item If $p=2$ the potential (\ref{poEin}) becomes constant for $\chi\gg\Mpl$.
We note that it is not necessary for $p$ to exactly equal $2$ for
this behaviour to occur. Since $U$ is approximated as $U\propto\chi^{2}$
in the regime $\chi\ll\Mpl$, inflation is followed by a successful
reheating. 
\item For $p>2$ the field rolls down the potential towards $\chi=0$. 
\item When $p<2$ the field rolls down the potential towards $\chi=+\infty$
or towards $\chi=0$. In the latter case the potential does not have 
a minimum at $\phi=0$.
As a result, reheating is problematic for $p<2$.
\end{itemize}
{}From Eqs.~(\ref{nRex}) and (\ref{rex}) it follows that 
\begin{eqnarray}
n_{{\rm s}}-1 & = & -\frac{\mu^{2}\left[4+2(3p-4)F+(p-2)^{2}F^{2}
\right]}{(F-1)^{2}}\,,\label{nRBD}\\
r & = & \frac{8\mu^{2}[2+(p-2)F]^{2}}{(F-1)^{2}}\,.\label{rBD}
\end{eqnarray}
The number of e-foldings (\ref{efold2}) reads 
\begin{eqnarray}
N & = & \frac{1}{2\mu^{2}}\left(F-F_{f}\right)+\frac{1}{2}\left(1-\frac{1}{\mu^{2}}\right)
\ln\left(\frac{F}{F_{f}}\right)\qquad(p=2)\,,\label{N1}\\
N & = & \frac{p}{2\mu^{2}(p-2)}\ln\left(\frac{2+(p-2)F}{2+(p-2)F_{f}}\right)+\frac{1}{2}\left(1-\frac{1}{\mu^{2}}\right)\ln\left(\frac{F}{F_{f}}\right)\qquad(p\ne2),\label{N2}
\end{eqnarray}
where $F_{f}$ is the value of $F$ at the end of inflation. Using
the criterion $\hat{\epsilon}=1$ for the end of inflation, we have
\begin{equation}
F_{f}=\frac{1+\sqrt{2}\mu}{1-(p-2)\mu/\sqrt{2}}\,.
\end{equation}
\subsubsection{Case: $p=2$}

Let us consider the case $p=2$. 
For the theories with $|\omega_{{\rm BD}}|\sim{\cal O}(1)$
(i.e. $\mu\sim{\cal O}(1)$) one has $F_{f}=1+\sqrt{2}\mu={\cal O}(1)$
and $N\simeq F/(2\mu^{2})$, which means that $F\gg1$ for $N\gg1$.
{}From Eqs.~(\ref{nRBD}) and (\ref{rBD}) it follows that 
\begin{eqnarray}
n_{{\rm s}}-1 & \simeq & -\frac{4\mu^{2}}{F}\simeq-\frac{2}{N}\,,\\
r & \simeq & \frac{32\mu^{2}}{F^{2}}\simeq\frac{8}{\mu^{2}N^{2}}=\frac{4(3+2\omega_{{\rm BD}})}{N^{2}}\,,
\end{eqnarray}
 which are valid for $-3/2<\omega_{{\rm BD}}<{\cal O}(1)$. The metric
$f(R)$ gravity corresponds to $\omega_{{\rm BD}}=0$, which gives
$r\simeq12/N^{2}$. This result matches with the one derived in other
papers \cite{fRspe}. 
In the limit that $\omega_{{\rm BD}}\to-3/2$
the tensor-to-scalar ratio vanishes. 
The BD parameter $\omega_{{\rm BD}}=-3/2$
corresponds to Palatini $f(R)$ gravity \cite{Sotiriou,DT10}, 
in which case a separate
analysis is required as in Ref.~\cite{Contaldi}.

If $\omega_{{\rm BD}}\gg1$, then one has $\mu\ll1$ and hence $F$
is close to 1 even during inflation. The end of inflation is characterized
by the condition $\hat{\epsilon}=1$, which gives $F_{f}=1+\sqrt{2}\mu\simeq1$.
Then the number of e-foldings (\ref{N1}) is approximately given by
$N\simeq(\chi/\Mpl)^{2}/4$. 
The scalar spectral index and the tensor-to-scalar ratio are
\begin{eqnarray}
n_{{\rm s}}-1 & \simeq & -8\frac{\Mpl^{2}}{\chi^{2}}\simeq-\frac{2}{N}\,,\\
r & \simeq & 32\frac{\Mpl^{2}}{\chi^{2}}\simeq\frac{8}{N}\,,
\end{eqnarray}
which match with those for the chaotic inflation model with the potential
$U(\phi)=m^{2}\phi^{2}/2$ \cite{Stewart}.

%%%%%%%%%%%%%%%%%%%%%%%%%%%%%
\begin{figure}
\begin{centering}
\includegraphics[width=5.5in,height=4in]{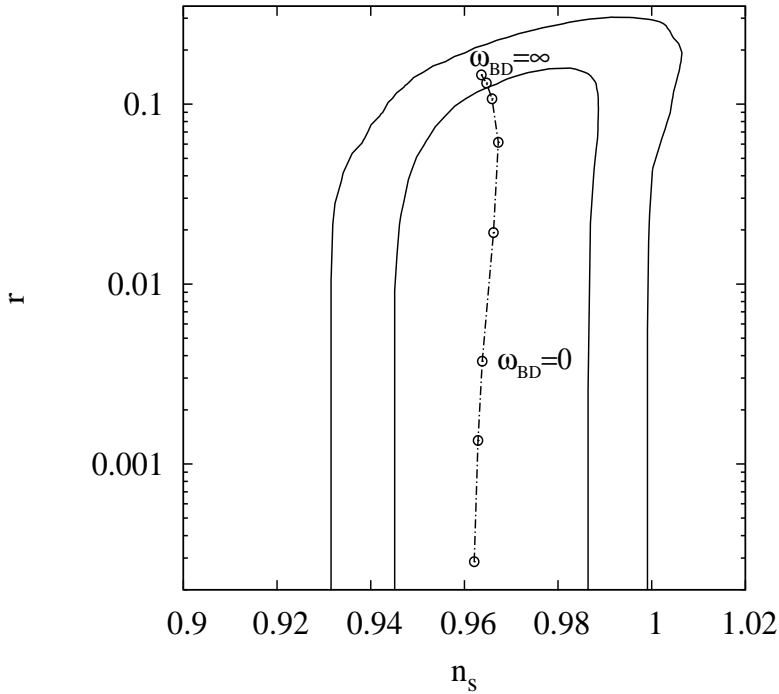} 
\par\end{centering}
\caption{1$\sigma$ and 2$\sigma$ observational contours in the $(n_{{\rm s}},r)$
plane constrained by the joint data analysis of WMAP 7yr, BAO, and
HST with the pivot scale $k_{0}=0.002$ Mpc$^{-1}$
(logarithmic scale for the vertical line). The dotted points show
the theoretical predictions for the BD theories with the potential
$V(\phi)=V_{0}(\phi-\Mpl)^{2}$. The number of e-foldings is chosen
to be $N=55$. {}
From bottom to top the points correspond to 
$\omega_{{\rm BD}}=-1.4, -1, 0, 10, 10^2, 10^3, 10^4$ and $\omega_{{\rm BD}}\to\infty$,
where $\omega_{{\rm BD}}=0$ represents the model $f(R)=R+R^{2}/(6M^{2})$.
For larger $\omega_{{\rm BD}}$ the two observables $n_{{\rm s}}$
and $r$ approach those for the chaotic inflation with the quadratic 
potential $m^{2}\phi^{2}/2$.}
\centering{}\label{fig3} 
\end{figure}
%%%%%%%%%%%%%%%%%%%%%%%%%%%%

The tensor-to-scalar ratio depends on the BD parameter $\omega_{{\rm BD}}$,
while the scalar index is practically independent of $\omega_{{\rm BD}}$.
In Fig.~\ref{fig3} we plot the theoretical predictions of $n_{{\rm s}}$
and $r$ for several different values of $\omega_{{\rm BD}}$ by fixing $N=55$.
Shown also are the 1$\sigma$ and 2$\sigma$ observational contours
constrained by the joint data analysis of WMAP 7yr \cite{WMAP7}, 
BAO \cite{BAO}, and HST \cite{HST}.
The $f(R)$ model $f(R)=R+R^{2}/(6M^{2})$, which corresponds to 
$\omega_{{\rm BD}}=0$, is well within the 1$\sigma$ 
observational contour. While the present
observations allow the large BD parameter
with $\omega_{{\rm BD}}\gg1$,
it will be of interest to see how the PLANCK satellite \cite{PLANCK}
can provide an upper bound on $\omega_{{\rm BD}}$.

Using the approximate relation $F \simeq 2\mu^2 N+F_f$ following from 
Eq.~(\ref{N1}), the WMAP normalization for the scalar power spectrum
${\cal P}_{\rm s}=U^3/(12 \pi^2 \Mpl^6 U_{,\chi}^2)$ is given by 
\begin{equation}
{\cal P}_{\rm s} \simeq \frac{V_0}{12 \pi^2 \Mpl^2}
\frac{(\sqrt{2}\mu N+1)^4 \mu^2}
{(2\mu^2 N+1+\sqrt{2} \mu)^2} =2.4 \times 10^{-9}\,,
\end{equation}
around $N=55$. 
Since $\omega_{\rm BD}=0$ and $\mu=1/\sqrt{3/2}$ for the $f(R)$ 
model $f(R)=R+R^2/(6M^2)$, the mass scale $M$ is constrained 
to be $M \simeq 3 \times 10^{13}$\,GeV.
The energy scale $V_0$ is different depending on 
the BD parameter.

\subsubsection{Case: $p\ne2$}

We proceed to the case $p\ne2$. For the BD parameter $\omega_{{\rm BD}}$
of the order of unity, the number of e-foldings (\ref{N2}) cannot
be much greater than 1 unless $F$ is enormously larger than 
$F_{f}~(\sim{\cal O}(1))$.
If $F\gg1$, then Eqs.~(\ref{nRBD}) and (\ref{rBD}) give 
\begin{eqnarray}
n_{\rm s}-1 & \simeq & -\mu^{2}(p-2)^{2}\,,
\label{nsrneq1} \\
r & \simeq & 8\mu^{2}(p-2)^{2}\,.
\label{nsrneq2}
\end{eqnarray}
Since $\mu\sim{\cal O}(1)$ the results (\ref{nsrneq1}) and 
(\ref{nsrneq2}) mean that for small $\wbd$ both
the scalar index and the tensor-to-scalar ratio are incompatible
with observations apart from the case where $p$ is close to 2. 
On reflection this is unsurprising since only for
$p\approx2$ is there a flat region of the potential that will give
slow-roll along with its signatures of near scale invariance 
and suppressed tensor modes.

When $\omega_{{\rm BD}}\gg1$ one has $\mu\ll1$ and 
hence $F=e^{\mu\chi/\Mpl}$ is close to 1. 
Then Eqs.~(\ref{nRBD}) and (\ref{rBD}) give 
\begin{eqnarray}
n_{\rm s}-1 & \simeq & -p(p+2)\frac{\Mpl^{2}}{\chi^{2}}\simeq-\frac{p+2}{2N}\,,\\
r & \simeq & 8p^{2}\frac{\Mpl^{2}}{\chi^{2}}\simeq\frac{4p}{N}\,,
\end{eqnarray}
where we have used the approximate relation $N\simeq\chi^{2}/(2p\Mpl^{2})$.
These results match with those of chaotic inflation with the
potential (\ref{chaoticpo}). The case $p=4$ is excluded
observationally both in the regimes $\omega_{{\rm BD}}\gg1$ 
and $\omega_{{\rm BD}}={\cal O}(1)$.
Even for other values of $\omega_{{\rm BD}}$ it is difficult to satisfy
observational constraints unless $p$ is close to 2.

%%%%%%%%%%%%%%%%%%%%%%%%%%%%%%%%%
\section{Inflation in the presence of a Gauss-Bonnet term}
\label{GBsec} 
%%%%%%%%%%%%%%%%%%%%%%%%%%%%%%%%%

In this section we study the effects of the Gauss-Bonnet (GB) term
on the chaotic inflationary scenario, described by the action 
\begin{equation}
S=\int d^{4}x\sqrt{-g}\left[\frac{M_{{\rm pl}}^{2}}{2}R
+X-V(\phi)-\xi(\phi){\cal G}\right]\,.
\end{equation}
In order to confront the model with observations, it is convenient
to rewrite inflationary observables in terms of the following 
slow-roll parameters: 
\begin{equation}
\epsilon_{\V}\equiv\frac{\Mpl^{2}}{2}\left(\frac{V_{,\phi}}
{V}\right)^{2}\,,\qquad\eta_{\V}\equiv\frac{\Mpl^{2}V_{,\phi\phi}}{V}\,.\label{epVetaV}
\end{equation}
The background equations are 
\begin{eqnarray}
 &  & 3\Mpl^{2}H^{2}=\dot{\phi}^{2}/2+V+24H^{3}\dot{\xi}\,,\label{sta1}\\
 &  & \ddot{\phi}+3H\dot{\phi}+V_{,\phi}+24H^{2}\xi_{,\phi}(H^{2}+\dot{H})=0\,.\label{sta2}
\end{eqnarray}
At linear order Eqs.~(\ref{eps}) and (\ref{epap}) 
give 
\begin{equation}
\epsilon_{s}=\delta_{X}\,,\qquad\epsilon=\epsilon_{s}+4\delta_{\xi}\,.
\end{equation}

{} From Eqs.~(\ref{sta1}) and (\ref{sta2}) the potential $V$ and
its derivative $V_{,\phi}$ can be expressed as 
\begin{eqnarray}
V & = & 3\Mpl^{2}H^{2}\left(1-\frac{1}{3}\epsilon_{s}-8\delta_{\xi}\right)\,,
\label{Veq}\\
V_{,\phi} & = & -H\dot{\phi}\left[3-\epsilon+\frac{1}{2}\eta_{s}+12\frac{\delta_{\xi}}{\epsilon_{s}}(1-\epsilon)\right]\,.\label{Vphi}
\end{eqnarray}
Taking the leading-order contribution in Eq.~(\ref{Vphi}), 
it follows that 
\begin{eqnarray}
V_{,\phi} & \simeq & -3H\dot{\phi}\left(1+\frac{4\delta_{\xi}}{\epsilon_{s}}\right)\,,\label{dVphi}\\
V_{,\phi\phi} & \simeq & -3H^{2}\left[\frac{1}{2}\eta_{s}-2\epsilon_{s}-16\delta_{\xi}-\frac{4\delta_{\xi}}{\epsilon_{s}}\left(8\delta_{\xi}+\frac{1}{2}\eta_{s}-\eta_{\xi}\right)\right]\,,
\end{eqnarray}
which lead to 
\begin{eqnarray}
\epsilon_{\V} & \simeq & \epsilon_{s}\left(1+\frac{4\delta_{\xi}}{\epsilon_{s}}\right)^{2}\,,\label{epVGB}\\
\eta_{\V} & \simeq & -\frac{1}{2}\eta_{s}\left(1-\frac{4\delta_{\xi}}{\epsilon_{s}}\right)+2\epsilon_{s}+4\delta_{\xi}\left[4+\frac{1}{\epsilon_{s}}(8\delta_{\xi}-\eta_{\xi})\right]\,.
\end{eqnarray}
{}From this we obtain the inversion formulas
\begin{eqnarray}
\epsilon_{s} & \simeq & \frac{1}{2}\left[\epsilon_{\V}-8\delta_{\xi}+\sqrt{\epsilon_{\V}^{2}-16\epsilon_{\V}\delta_{\xi}}\right]\,,\label{epsin}\\
\eta_{s} & \simeq & -\frac{2\left\{ \eta_{\V}-2\epsilon_{s}-4\delta_{\xi}[4+(8\delta_{\xi}-\eta_{\xi})/\epsilon_{s}]\right\} }{1-4\delta_{\xi}/\epsilon_{s}}\,,\label{etasGB}
\end{eqnarray}
where we have taken the positive sign in Eq.~(\ref{epsin}) to reproduce
$\epsilon_{s} \to \epsilon_{\V}$ for $\delta_{\xi} \to 0$.

{}From Eqs.~(\ref{nR}), (\ref{nT}), and (\ref{rgene}) the inflationary
observables are given by 
\begin{eqnarray}
n_{{\rm s}}-1 & = & -2\epsilon_{s}-\eta_{s}-8\delta_{\xi}\,,\label{nsGB0}\\
n_{{\rm t}} & = & -2\epsilon_{s}-8\delta_{\xi}\,,\label{ntGB0}\\
r & = & 16\epsilon_{s}\,,\label{rGB0}
\end{eqnarray}
which are written in terms of the four variables: $\epsilon_{\V}$,
$\eta_{\V}$, $\delta_{\xi}$, and $\eta_{\xi}$. By specifying the
functional forms of $V(\phi)$ and $\xi(\phi)$, we can reduce the
number of those variables. For the chaotic inflation potential (\ref{chaoticpo})
one has $\epsilon_{\V}=(p^{2}/2)(\Mpl/\phi)^{2}$ and 
$\eta_{\V}=p(p-1)(\Mpl/\phi)^{2}$,
so that they are related with each other via the relation 
\begin{equation}
\eta_{\V}=\frac{2(p-1)}{p}\epsilon_{\V}\,.
\label{etaepsilonpowerlaw}
\end{equation}
For the Gauss-Bonnet coupling, we take
\begin{equation}
\xi(\phi)=\xi_{0}e^{\mu\phi/\Mpl}\,,
\end{equation}
where $\xi_{0}$ and $\mu$ are constants. 
It then follows that 
\begin{equation}
\eta_{\xi}=-2\epsilon_{s}+\eta_{s}/2-8\delta_{\xi}\pm\mu\sqrt{2\epsilon_{s}}\,,\label{etaxi}
\end{equation}
where the plus and minus signs correspond to $\dot{\phi}>0$ and
$\dot{\phi}<0$, respectively. Combining Eq.~(\ref{etaxi}) with
Eq.~(\ref{etasGB}), we obtain 
\begin{equation}
\eta_{s}\simeq-2\left[\eta_{\V}-2\epsilon_{s}+4\delta_{\xi}\left(\frac{\pm\mu\sqrt{2\epsilon_{s}}-16\delta_{\xi}}{\epsilon_{s}}-6\right)\right]\,.\label{etasap}
\end{equation}
Substituting Eq.~(\ref{etasap}) into Eq.~(\ref{nsGB0}) and choosing
the negative sign of $\dot{\phi}$, the scalar spectral index can
be written as 
\begin{eqnarray}
n_{{\rm s}}-1 & \simeq & -6\epsilon_{s}+2\eta_{\V}-8\delta_{\xi}\left(7+\frac{\mu\sqrt{2\epsilon_{s}}+16\delta_{\xi}}{\epsilon_{s}}\right)\,,\label{nsGB}
\end{eqnarray}
where 
\begin{equation}
\eta_{\V}=\frac{2(p-1)}{p}\epsilon_{\V}\simeq\frac{2(p-1)}{p}\epsilon_{s}\left(1+\frac{4\delta_{\xi}}{\epsilon_{s}}\right)^{2}\,.\label{etaV}
\end{equation}
For fixed values of $p$ and $\mu$ one can carry out the CMB likelihood
analysis in terms of $n_{\rm s}$, $r$, and $n_{\rm t}$
by varying the two parameters $\epsilon_{s}$ and $\delta_{\xi}$.

\begin{figure}
\begin{centering}
\includegraphics[width=3.6in,height=3.4in]{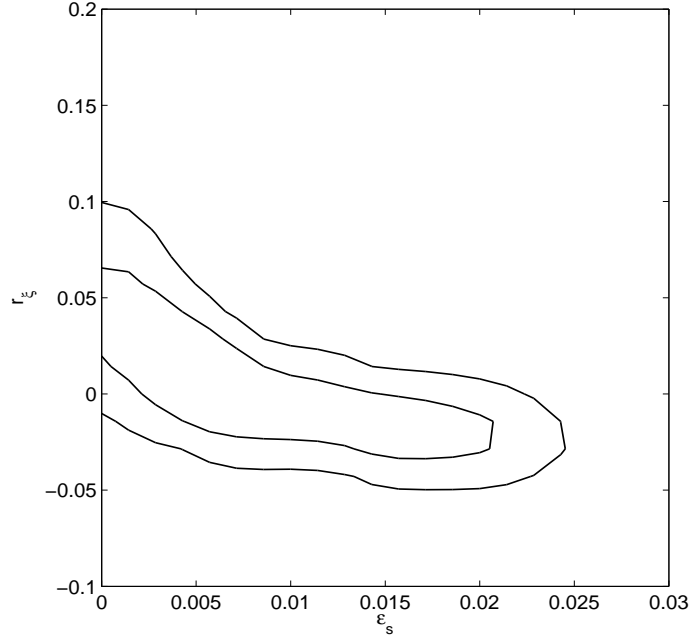} 
\par\end{centering}
\caption{1$\sigma$ (inside) and $2\sigma$ (outside) observational contours
in the $(\epsilon_{{\rm s}},r_{\xi})$ plane for the potential $V(\phi)=m^{2}\phi^{2}/2$
with $\mu=1$ ($r_{\xi}=\delta_{\xi}/\epsilon_{s}$). We use the data
of WMAP 7 yr, LSS (including BAO), HST, SN Ia, and BBN with the pivot
scale $k_{0}=0.002$ Mpc$^{-1}$.}
\centering{}\label{fig4} 
\end{figure}

In Fig.~\ref{fig4} the observational constraints on the parameters
$\epsilon_{s}$ and $r_{\xi}\equiv\delta_{\xi}/\epsilon_{s}$ are
plotted for $p=2$ and $\mu=1$. We run the Cosmological Monte Carlo
(CosmoMC) code \cite{cosmomc} with the data of WMAP 7yr \cite{WMAP7}
combined with large-scale structure \cite{Reid} (including BAO \cite{BAO}),
HST \cite{HST}, Supernovae type Ia (SN Ia) \cite{SNIa}, and Big
Bang Nucleosynthesis (BBN) \cite{BBN}, by assuming a $\Lambda$CDM
universe. The ratio $r_{\xi}$ is constrained to be $|r_{\xi}|<0.1$ (95\%~{\rm CL}),
which means that the effect of the GB term needs to be suppressed.
Hence the energy scale $V_0$ is similar to that in the standard 
chaotic inflation.

{}From Fig.~\ref{fig4} we find that the slow-roll parameter 
$\epsilon_{s}$ is bounded to be $\epsilon_{s}<0.025$
(95 \% CL). In the presence of the GB term the small values of $\epsilon_{s}$
can give rise to the scalar index close to $n_{s}=0.96$. For example,
when $\epsilon_{s}=0.002$, $r_{\xi}=0.05$, $\mu=1$, and $p=2$,
one has $n_{{\rm s}}=0.962$ from Eq.~(\ref{nsGB}). This is different
from the standard chaotic inflation in which the small values of $\epsilon_{s}$
lead to the spectrum close to the Harrison-Zel'dovich one (which
is not favored observationally). Hence the allowed range of $\epsilon_{s}$
tends to be wider in the presence of the GB coupling.

%%%%%%%%%%%%%%%%%%%%%%%%%%%%%
%%%%%%%%%%%%%%%%%%%%%%%%%%%%

Let us estimate the two observables $n_{{\rm s}}$ and $r$ in terms
of the number of e-foldings $N$ under the condition $|r_{\xi}|\ll1$.
Since $\epsilon_{s}\simeq\epsilon_{V}-8\delta_{\xi}$ from Eq.~(\ref{epsin}),
Eqs.~(\ref{nsGB}) and (\ref{rGB0}) reduce to 
\begin{eqnarray}
n_{{\rm s}}-1 & \simeq & -\left(2+\frac{4}{p}\right)\epsilon_{\V}-8\delta_{\xi}\left(1+\mu\sqrt{\frac{2}{\epsilon_{\V}}}\right)\,,\label{nsGB2}\\
r & \simeq & 16\epsilon_{\V}\left(1-\frac{8\delta_{\xi}}{\epsilon_{\V}}\right)\,.\label{rGB2}
\end{eqnarray}
{}From Eqs.~(\ref{Veq}) and (\ref{dVphi}) one has $H/\dot{\phi}\simeq-(1+4\delta_{\xi}/\epsilon_{s})V/(\Mpl^{2}V_{,\phi})$.
Using the relation $\epsilon_{s}=\delta_{X}$ and 
the definition of $\delta_{\xi}$, 
it follows that $H/\dot{\phi}=-\phi/(p\Mpl^{2}+8H^{2}\xi_{,\phi}\phi)$.
Since we are considering the case where $H^{2}\xi_{,\phi}\phi/\Mpl^{2}\ll1$,
the number of e-foldings for the potential (\ref{chaoticpo}) is 
\begin{equation}
N=\int_{\phi}^{\phi_{f}}\frac{H}{\dot{\phi}}\, d\phi\simeq\frac{x^{2}-x_{f}^{2}}{2p}+\bar{N}_{p}\,,\qquad{\rm where}\qquad\bar{N}_{p}\equiv-\frac{8\xi_{0}\mu}{3p^{2}}\frac{V_{0}}{\Mpl^{4}}\int_{x_{f}}^{x}e^{\mu x}x^{p+2}dx\,.\label{efoldap}
\end{equation}
Here $x_{f}$ is the value of $x=\phi/\Mpl$ at the end of inflation.
We identify the end of inflation by the condition $\epsilon_{\V}=1$,
i.e. $x_{f}=p/\sqrt{2}$. When $p=2$ and $p=4$, Eq.~(\ref{efoldap})
is integrated to give 
\begin{eqnarray}
\bar{N}_{2} & = & -\frac{m^{2}}{3\Mpl^{2}}\frac{\xi_{0}}{\mu^{4}}\left\{ e^{\mu x}[\mu^{4}x^{4}+4(-\mu^{3}x^{3}+3\mu^{2}x^{2}-6\mu x+6)]-4e^{\sqrt{2}\mu}\,[\mu^{4}-2\sqrt{2}\mu^{3}+6(\mu^{2}-\sqrt{2}\mu+1)]\right\} \,,\label{N2d}\\
\bar{N}_{4} & = & -\frac{\lambda\xi_{0}}{24\mu^{6}}\,\biggl\{ e^{\mu x}(\mu^{6}x^{6}-6\mu^{5}x^{5}+30\mu^{4}x^{4}-120\mu^{3}x^{3}+360\mu^{2}x^{2}-720\mu x+720)\nonumber \\
 &  & ~~~~~~~~~~\,-16e^{2\sqrt{2}\mu}\left[32\mu^{6}-48\sqrt{2}\mu^{5}+120\mu^{3}(\mu-\sqrt{2})+90\mu(2\mu-\sqrt{2})+45\right]\biggr\}\,,\label{N4d}
\end{eqnarray}
where we have set $V_{0}=m^{2}\Mpl^{2}/2$ for $p=2$ and 
$V_{0}=\lambda\Mpl^{4}/4$ for $p=4$.

For positive $\mu$ of the order of unity, the dominant contributions
to $\bar{N}_{p}$ come from the first terms in Eqs.~(\ref{N2d})
and (\ref{N4d}), i.e. 
$\bar{N}_{2}\simeq-m^{2}\xi_{0} x^{4} e^{\mu x}
/(3\Mpl^{2})\,$
and $\bar{N}_{4}\simeq-\lambda\xi_{0} x^{6} e^{\mu x}/24$, 
for the scales relevant to CMB ($\mu x\gg1$). In this case the number
of e-foldings (\ref{efoldap}) is approximately given by 
\begin{eqnarray}
N & \simeq & \frac{1}{4}x^{2}\left[1-\frac{4}{3}\left(\frac{m}{\Mpl}\right)^{2}\xi x^{2}\right]-\frac{1}{2}\qquad(p=2)\,,\label{Np2}\\
N & \simeq & \frac{1}{8}x^{2}\left(1-\frac{1}{3}\lambda\xi x^{4}\right)-1\qquad(p=4)\,.\label{Np4}
\end{eqnarray}
Since $\delta_{\xi}\simeq-\mu V_{0}p/(3\Mpl^{4})\, x^{p-1}\xi$ and
$\epsilon_{\V}=p^{2}/(2x^{2})$, one can express the scalar index
(\ref{nsGB2}) and the tensor-to-scalar ratio (\ref{rGB2}) in terms of 
 $x$. By treating the $\xi$-dependent terms in Eqs.~(\ref{Np2})
and (\ref{Np4}) as small corrections, $n_{{\rm s}}$ and $r$ can
be written in terms of $N$: 
\begin{eqnarray}
n_{{\rm s}}-1 & \simeq & -\frac{2}{N}\left[1-\frac{16}{3}N^{2}\left(\frac{m}{\Mpl}\right)^{2}\mu^{2}\xi\right]\,,\qquad r\simeq\frac{8}{N}\left[1+\frac{32}{3}N^{3/2}\left(\frac{m}{\Mpl}\right)^{2}\mu\,\xi\right]\,,\qquad(p=2)\,,\\
n_{{\rm s}}-1 & \simeq & -\frac{3}{N}\left[1-\frac{256}{9}N^{3}\lambda\mu^{2}\xi\right]\,,\qquad~~~~~~~~~\, r\simeq\frac{16}{N}\left[1+\frac{128\sqrt{2}}{3}N^{5/2}\lambda\mu\,\xi\right]\,,\qquad~~~~~(p=4)\,,
\end{eqnarray}
which are valid for positive $\mu$ of the order of unity. If $\xi>0$
(i.e. $\delta_{\xi}<0$ for $\dot{\phi}<0$), then the effect of the
GB coupling leads to the approach to the scale-invariant spectrum,
while $r$ gets larger. The negative values of $\xi$ lead to the
decease of $r$, but $n_{{\rm s}}$ deviates from 1. Since $\epsilon_{s}$
is approximately given by $\epsilon_{s}\approx\epsilon_{\V}\approx p^{2}/(8N)$,
the scales relevant to the CMB anisotropies ($N=50$-60) correspond
to $0.008<\epsilon_{s}<0.01$ for $p=2$. Figure \ref{fig4} shows
that the ratio $r_{\xi}$ is constrained to be $-0.04<r_{\xi}<0.03$
(95 \% CL) for this range of $\epsilon_{s}$. The self-coupling potential
$V(\phi)=\lambda\phi^{4}/4$ is not saved by taking into account the
GB term with positive $\mu$, because the GB coupling does not lead
to the increase of $n_{{\rm s}}$ and the decrease of $r$ simultaneously.

For negative $\mu$ with $|\mu|={\cal O}(1)$ the exponential term
$e^{\mu x}$ in Eqs.~(\ref{N2d}) and (\ref{N4d}) is much smaller
than 1 for the scales relevant to CMB ($x\gg1$).
In this case we have 
\begin{eqnarray}
\bar{N}_{2} & \simeq & \frac{4m^{2}}{3\Mpl^{2}}\frac{\xi_{0}e^{\sqrt{2}\mu}}{\mu^{4}}\left[\mu^{4}-2\sqrt{2}\mu^{3}+6(\mu^{2}-\sqrt{2}\mu+1)\right]\,,\label{N2ne}\\
\bar{N}_{4} & \simeq & \frac{2\lambda\xi_{0}e^{2\sqrt{2}\mu}}{3\mu^{6}}\left[32\mu^{6}-48\sqrt{2}\mu^{5}+120\mu^{3}(\mu-\sqrt{2})+90\mu(2\mu-\sqrt{2})+45\right]\,.\label{N4ne}
\end{eqnarray}
The scalar index and the tensor-to-scalar ratio are approximately given by 
\begin{eqnarray}
n_{{\rm s}}-1 & \simeq & -\frac{2}{N}\left(1+\frac{\bar{N}_{2}-1/2}{N}\right),\qquad r\simeq\frac{8}{N}\left(1+\frac{\bar{N}_{2}-1/2}{N}\right),\qquad(p=2),\\
n_{{\rm s}}-1 & \simeq & -\frac{3}{N}\left(1+\frac{\bar{N}_{4}-1}{N}\right),\qquad~~~r\simeq\frac{16}{N}\left(1+\frac{\bar{N}_{4}-1}{N}\right),\qquad~~~(p=4),
\end{eqnarray}
where we have assumed $N\gg\bar{N}_{p}$, and ignored the exponential
term $e^{\mu x}$. When $\mu<0$, one can show that $\bar{N}_{2}$
and $\bar{N}_{4}$ in Eqs.~(\ref{N2ne}) and (\ref{N4ne}) are positive
for $\xi_{0}>0$ and negative for $\xi_{0}<0$. In the latter case
the presence of the GB term leads to the approach to the Harrison
Zel'dovich spectrum.
In fact, such a scenario was discussed in Ref.~\cite{Satoh}.
Since $m/M_{{\rm pl}}$ and $\lambda$ are much smaller than 1 
by the WMAP normalization ($m/M_{{\rm pl}} \simeq 6.8 \times 10^{-6}$
and $\lambda \simeq 2.0 \times10^{-13}$), one has $|\bar{N}_{p}|\ll1$
for $|\xi_{0}|$ smaller than the order of 1.
For $\mu<0$ the effect of the GB term on the inflationary observables 
appears for very large values of $\xi_0$ such as 
$|\xi_0| \sim 10^{10}$ \cite{Satoh}.

%%%%%%%%%%%%%%%%%%%%%%%%
\section{G-inflation with a field potential}
\label{Galileonsec} 
%%%%%%%%%%%%%%%%%%%%%%%%

Finally we study chaotic inflation in the presence of the Galileon-like self-interaction
$G(\phi,X)\square\phi$ (called ``G-inflation''). 
We specify the functional form of $G(\phi,X)$, as 
\begin{equation}
G(\phi,X)=\Phi(\phi)X^{n}\,,\qquad\Phi(\phi)=
\frac{\theta}{M^{4n-1}}e^{\mu\phi/\Mpl}\,,\label{Phiex}
\end{equation}
where $n$ and $\mu$ are constants, and $\theta=\pm1$. The constant
$M$ has a dimension of mass with $M>0$. Here we have introduced
the exponential form for $\Phi$ motivated by the dilaton coupling
in the low-energy effective bosonic string theory. We also consider
the power-law function $X^{n}$ by generalizing previous studies \cite{Kamada-etal-10}.

Equations (\ref{E1eq}) and (\ref{E3eq}) can be written as 
\begin{eqnarray}
&  & V=3\Mpl^{2}H^{2}\left(1-\frac{1}{3}\delta_{X}-2\delta_{GX}
+\frac{2}{3}\delta_{G\phi}\right)\,,\label{bega1}\\
&  & V_{,\phi}=-3H\dot{\phi}\left\{ 1+(3-\epsilon)
\frac{\delta_{GX}}{\delta_{X}}-\frac{\mu^{2}}{3n}\delta_{GX}
+2(n-1)\frac{\delta_{G\phi}}{\delta_{X}}+\frac{\delta_{\phi}}{3}
\left[1+6n\frac{\delta_{GX}}{\delta_{X}}-2(n+1)
\frac{\delta_{G\phi}}{\delta_{X}}\right]\right\} \,.\label{bega2}
\end{eqnarray}
To compare with observations, we seek an expression for 
$n_{{\rm s}}-1=\dot{\mathcal{P}}_{{\rm s}}/(H\mathcal{P}_{{\rm s}})$
in terms of a minimal set of independent slow-roll parameters. Since
$\mathcal{P}_{{\rm s}}=H^{2}/(8\pi^{2}Qc_{s}^{3})$, it is important to
find the expressions for $Q$ and $c_{s}$. {}From Eqs.~(\ref{eq:defQ})
and (\ref{eq:defc2s}) it follows that 
\begin{eqnarray}
\frac{Q}{\Mpl^{2}} & = & \frac{\delta_{X}+6n\delta_{GX}-2(n+1)\delta_{G\phi}+3\delta_{GX}^{2}}{(1-\delta_{GX})^{2}}\,,\label{eq:QG1}\\
c_{s}^{2} & = & \frac{\delta_{X}+2(2+n\delta_{\phi})\delta_{GX}+2(n-1)\delta_{G\phi}-\delta_{GX}^{2}}{\delta_{X}+6n\delta_{GX}-2(n+1)\delta_{G\phi}+3\delta_{GX}^{2}}\,,\label{eq:cG1}
\end{eqnarray}
where we have used the relations 
$\lambda_{GX}=n-1$, and $\lambda_{G\phi}=n$.
Hence, we can derive an exact expression for $n_{{\rm s}}-1$ in terms
of the slow-roll parameters entering Eqs.~(\ref{eq:QG1}) and (\ref{eq:cG1})
and their first derivatives, which introduce other slow-roll
parameters, however these are not all independent. 
We shall now discuss the relations which reduce the
number of independent slow-roll parameters.

For the choice of the function $\Phi$ in Eq.~(\ref{Phiex}) we have
\begin{equation}
\delta_{G\phi}=\pm\frac{\mu}{\sqrt{2}n}\,\delta_{GX}\sqrt{\delta_{X}}\,,\label{eq:dGX1}
\end{equation}
where the $\pm$ signs in this expression are compatible with those in the expression
$\dot{\phi}=\pm\sqrt{2}\Mpl H\sqrt{\delta_{X}}$.
Equation (\ref{eq:dGX1}) shows that $\delta_{G\phi}$ is in general
suppressed relative to $\delta_{GX}$. This relation also implies
that 
\begin{equation}
\eta_{G\phi}\equiv\frac{\dot{\delta}_{G\phi}}{H\delta_{G\phi}}
=\frac{\eta_{X}}{2}+\eta_{GX}\,,\label{etaGphi}
\end{equation}
where 
\begin{equation}
\eta_{X}\equiv\frac{\dot{\delta}_{X}}{H\delta_{X}}\,,\qquad{\rm and}
\qquad\eta_{GX}=\frac{\dot{\delta}_{GX}}{H\delta_{GX}}\,.
\end{equation}
 From the definition of $\delta_{X}$ and $\delta_{GX}$ we obtain
\begin{eqnarray}
\eta_{X} & = & 2(1-\delta_{GX})\delta_{\phi}+2\delta_{X}+6\delta_{GX}
-4\delta_{G\phi}\,,\label{etaX}\\
\eta_{GX} & = & (2n+1-\delta_{GX})\delta_{\phi}\pm\mu\sqrt{2\delta_{X}}
+\delta_{X}-2\delta_{G\phi}+3\delta_{GX}\,,\label{eq:eGX1}
\end{eqnarray}
 where we have used the relation 
\begin{equation}
\epsilon=\delta_{X}+3\delta_{GX}-2\delta_{G\phi}-\delta_{\phi}\delta_{GX}\,.\label{epGa}
\end{equation}
 It should be noted that the four relations~(\ref{eq:dGX1})-(\ref{epGa})
are all exact. {}From Eqs.~(\ref{etaGphi}), (\ref{etaX}), (\ref{eq:eGX1})
with Eq.~(\ref{eq:dGX1}) we find that $\eta_{G\phi}$, 
$\eta_{X}$,
and $\eta_{GX}$ can be expressed in terms of the three slow-roll
parameters $\delta_{\phi}$, $\delta_{X}$, and $\delta_{GX}$.

We finally use a last constraint coming from 
the fact that we have chosen a power-law form 
(\ref{chaoticpo}) for the potential. The relation
(\ref{etaepsilonpowerlaw}) between $\epsilon_{V}$ and $\eta_{V}$
leads to 
\begin{equation}
\frac{\dot{V}_{,\phi}}{HV_{,\phi}}=\frac{p-1}{p}\frac{V_{,\phi}}{HV}\,\dot{\phi}\,.
\end{equation}
This equation can be used to set the last constraint on the slow-roll
variables. At lowest order we have 
\begin{equation}
\delta_{\phi}=\frac{(\delta_{X}+3\delta_{GX})[(2-p)\delta_{X}+6\delta_{GX}]}
{p(\delta_{X}+6n\delta_{GX})} \mp \frac{3\sqrt{2}\delta_{GX}\sqrt{\delta_{X}}}
{\delta_{X}+6n\delta_{GX}}\mu-\frac{2(n-1)\delta_{X}\delta_{GX}
(\delta_{X}-3\delta_{GX})}{n(\delta_{X}+6n\delta_{GX})^{2}}\mu^{2}
+{\cal O}(\epsilon^{3/2})\,.\label{eq:d2pF}
\end{equation}
Using this relation we can express $\eta_{G\phi}$, $\eta_{X}$,
and $\eta_{GX}$ in terms of two slow-roll parameters $\delta_{X}$
and $\delta_{GX}$.

We are now ready to explicitly calculate the scalar index $n_{{\rm s}}-1=-2\epsilon-\delta_{Q}-3s$,
where $\delta_{Q}=\dot{Q}/(HQ)$ and $s=\dot{c}_{s}/(Hc_{s})$ are
evaluated by taking the time derivatives of Eqs.~(\ref{eq:QG1})
and (\ref{eq:cG1}). This gives 
\begin{eqnarray}
n_{{\rm s}}-1 & \simeq & -\frac{2\left(\delta_{{X}}+3\,\delta_{{GX}}\right)}{p\left(\delta_{{X}}+4\,\delta_{{GX}}\right)\left(\delta_{{X}}+6n\,\delta_{{GX}}\right)^{2}}\bigl[{\delta_{{X}}^{3}}(p+2)+{\delta_{{X}}^{2}}\delta_{{GX}}[(3p-6)n^{2}+(12p+27)n+4p+8]\nonumber \\
 &  & {}+\delta_{{X}}{\delta_{{GX}}^{2}}[(57p+30)n^{2}+(54p+105)n+6]+72n{\delta_{{GX}}^{3}}(3np+2n+1)\bigr]\nonumber \\
 &  & {} \pm {\frac{3\sqrt{2}\,\delta_{{GX}}\sqrt{\delta_{{X}}}\,[(7n+2)\,\delta_{{X}}\delta_{{GX}}+n{\delta_{{X}}^{2}}+24n\,{\delta_{{GX}}^{2}}]}{\left(\delta_{{X}}+6n\,\delta_{{GX}}\right)^{2}\left(\delta_{{X}}+4\,\delta_{{GX}}\right)}}\mu\nonumber \\
 &  & {}-\frac{2\delta_{{\it GX}}\delta_{{X}}}{n\left(\delta_{{X}}+6\, n\delta_{{\it GX}}\right)^{3}\left(4\,\delta_{{\it GX}}+\delta_{{X}}\right)^{2}}\,\bigl[{\delta_{{X}}}^{4}+\left(9\, n+8+6\,{n}^{3}-24\,{n}^{2}\right){\delta_{{X}}^{3}}\delta_{{\it GX}}\nonumber \\
 &  & {}+\left(4-99\,{n}^{2}+54\, n-42\,{n}^{3}\right){\delta_{{X}}^{2}}{\delta_{{\it GX}}^{2}}+\left(132\,{n}^{2}-282\,{n}^{3}-24-132\, n\right)\delta_{{X}}{\delta_{{\it GX}}^{3}}-72n\left(n^{2}-3n+6\right){\delta_{{\it GX}}^{4}}\bigr]\mu^{2},\nonumber \\
\label{nscom}
\end{eqnarray}
where, in order to derive this result, we have also included the
terms of order $\mathcal{O}(\epsilon^{3/2})$ not shown
in Eq.~(\ref{eq:d2pF}).
Again the $\pm$ signs in the
term involving $\mu$ in Eq.~(\ref{nscom})
are compatible with those in the expression 
for $\dot{\phi}$ in terms of $\delta_{X}$.
The tensor-to-scalar ratio (\ref{rgene}) and the tensor index (\ref{nT})
are approximately given by 
\begin{eqnarray}
r & \simeq & 16\frac{(\delta_{X}+4\delta_{GX})^{3/2}}
{(\delta_{X}+6n\delta_{GX})^{1/2}}\,,\label{rcom}\\
n_{{\rm t}} & \simeq & -2(\delta_{X}+3\delta_{GX})\,,\label{ntcom}
\end{eqnarray}
 and the scalar propagation speed squared is 
\begin{equation}
c_{s}^{2}\simeq\frac{\delta_{X}+4\delta_{GX}}
{\delta_{X}+6n\delta_{GX}}\,.\label{scalarga}
\end{equation}

If $|\delta_{GX}|\ll\delta_{X}$, then these observables reduce to
\begin{eqnarray}
n_{{\rm s}}-1 & \simeq & -\frac{2(p+2)}{p}\delta_{X}
\pm 3\sqrt{2}n\mu\frac{\delta_{GX}}{\sqrt{\delta_{X}}}\,,
\label{nslimi1}\\
r & \simeq & 16\delta_{X}\simeq-8n_{{\rm t}}\,.\label{nsrga}
\end{eqnarray}
On the other hand, in the limit where $\delta_{GX}\gg\delta_{X}$,
one has 
\begin{eqnarray}
n_{{\rm s}}-1 & \simeq & -\frac{3(3np+2n+1)}{pn}\delta_{GX}
 \pm \frac{\mu}{\sqrt{2}n}\,
\sqrt{\delta_{X}}\,,
\label{nsrga0} \\
r & \simeq & \frac{64}{3}\sqrt{\frac{6}{n}}\delta_{GX}\simeq
-\frac{32}{9}\sqrt{\frac{6}{n}}n_{{\rm t}}\simeq
-\frac{8.7}{\sqrt{n}}\, n_{{\rm t}}\,,\label{nsgali}
\end{eqnarray}
which agree with the results in Ref.~\cite{Kamada-etal-10} derived
for $n=1$ and $\mu=0$.

To be concrete, in the following discussion we focus on the theories 
with $n=1$, $\mu\neq 0$, and $\theta=-1$. 
Then $\delta_{GX}>0$ for $\dot{\phi}<0$, 
so that the conditions for the avoidance of ghosts and Laplacian 
instabilities ($Q>0$ and $c_{s}^{2}>0$) are always satisfied. 
In this case we need to take the minus sign 
for the term $\mu$ in Eqs.~(\ref{nscom}), (\ref{nslimi1}), and 
(\ref{nsrga0}).
Since $V_{,\phi}\simeq-3H\dot{\phi}(1+3H\dot{\phi}\Phi)$
from Eq.~(\ref{bega2}), the field velocity corresponding 
to $\dot{\phi}<0$ is 
\begin{equation}
\dot{\phi}\simeq\frac{\sqrt{1-4\Phi V_{,\phi}}-1}{6H\Phi}\,,
\end{equation}
 where we used $\Phi<0$. 
Employing the approximate relation $V\simeq3H^{2}\Mpl^{2}$,
the two slow-roll parameters $\delta_{X}$ and $\delta_{GX}$ can
be expressed in terms of $\phi$, as 
\begin{equation}
\delta_{X}\simeq\frac{\Mpl^{2}(\sqrt{1-4\Phi V_{,\phi}}-1)^{2}}
{8V^{2}\Phi^{2}}\,,\qquad\delta_{GX}\simeq\frac{\delta_{X}}{6}
(\sqrt{1-4\Phi V_{,\phi}}-1)\,.\label{delXGX}
\end{equation}
The number of e-foldings is given by 
\begin{equation}
N=\int_{\phi}^{\phi_{f}}\frac{H}{\dot{\phi}}\, d\phi\simeq\frac{2}{\Mpl^{2}}
\int_{\phi}^{\phi_{f}}\frac{\Phi V}{\sqrt{1-4\Phi V_{,\phi}}-1}d\phi
=2B^{4}\int_{x_{f}}^{x}\frac{x^{p}e^{\mu x}}{\sqrt{1
+4B^{4}px^{p-1}e^{\mu x}}-1}dx\,,\label{efoldGa}
\end{equation}
where 
\begin{equation}
B\equiv\left(\frac{V_{0}}{M^{3}\Mpl}\right)^{1/4}\,,\qquad 
x\equiv\frac{\phi}{\Mpl}\,,\qquad x_{f}\equiv\frac{\phi_{f}}{\Mpl}\,.
\end{equation}
We determine the value of $x_{f}$ at the end of inflation using the
condition $\epsilon\simeq\delta_{X}+3\delta_{GX}=1$.

%%%%%%%%%%%%%%%%%%%%%%%%%%%%%
\begin{figure}
\begin{centering}
\includegraphics[width=4.8in,height=3.5in]{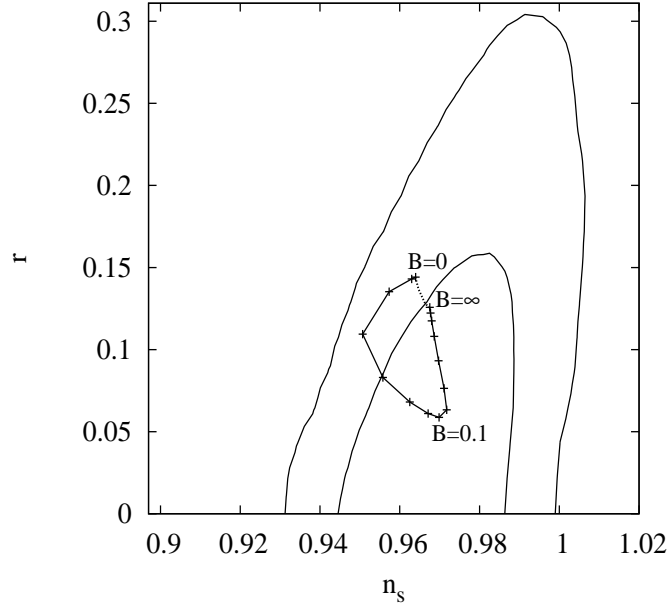} 
\par\end{centering}
\caption{Theoretical values of $n_{{\rm s}}$ and $r$ 
for the potential $V(\phi)=m^{2}\phi^{2}/2$
in the presence of the Galileon-type coupling $G=-(1/M^{3})e^{\mu\phi/\Mpl}X$
with $\mu=1$ (solid line).
The points correspond to the cases with
$B=0,10^{-5/2},10^{-9/4},10^{-2},10^{-7/4},10^{-3/2},10^{-5/4},0.1,
10^{-1/2},1,10^{1/2},10,10^{3/2},10^{2}$ with $N=55$.
In the limit $B \to \infty$ one has $n_{{\rm s}}=0.9675$
and $r=0.1258$. 
The dotted curve corresponds to the case where $\mu=0$.
We also show the $1\sigma$ and $2\sigma$ observational
contours derived by the joint data analysis of WMAP 7 yr, BAO, and
HST with the consistency relation $r=-8n_{{\rm t}}$.}
\centering{}\label{gap2} 
\end{figure}
%%%%%%%%%%%%%%%%%%%%%%%%%%%%

%%%%%%%%%%%%%%%%%%%%%%%%%%%%%
\begin{figure}
\begin{centering}
\includegraphics[width=4.8in,height=3.5in]{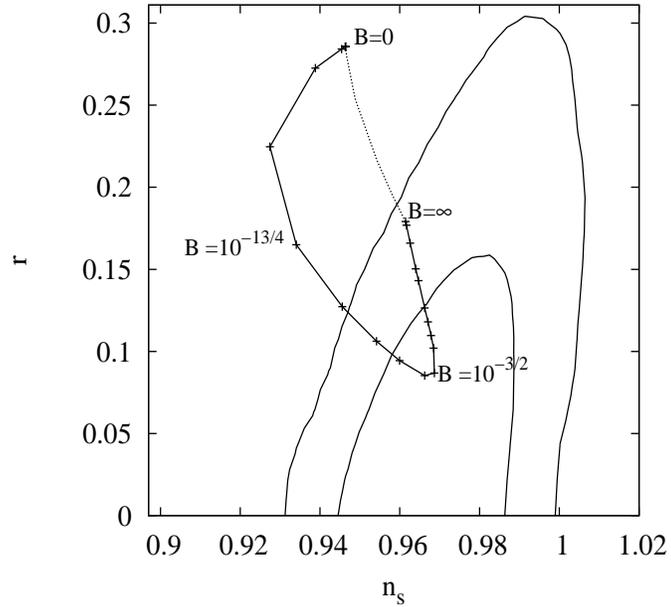} 
\par\end{centering}
\caption{Similar to Fig.~\ref{gap2}, but for the potential $V(\phi)=\lambda\phi^{4}/4$
with $\mu=1$ (solid line).
The points correspond to the cases with
$B=0,10^{-9/2},10^{-17/4},10^{-4},10^{-15/4},10^{-7/2},10^{-13/4},10^{-3},
10^{-11/4},10^{-5/2},10^{-2},10^{-3/2},10^{-3/4},10^{-1/2},10^{-1/4},1,
10^{1/2},10^{3/4}$, $10^{3/2},10^{3}$ with $N=55$.
In the limit where $B\to\infty$ one has $n_{{\rm s}}=0.9614$
and $r=0.1791$. 
The dotted curve corresponds to the case where $\mu=0$.
The same observational contours as those in Fig.~\ref{gap2}
are also plotted.}
\centering{}\label{gap4} 
\end{figure}
%%%%%%%%%%%%%%%%%%%%%%%%%%%%

In the limit $B \to 0$ (i.e. $\delta_{GX} \to 0$) we have
$\epsilon\simeq\delta_{X}\simeq p^{2}/(2x^{2})$ and 
$N\simeq x^{2}/(2p)-p/4$, so that
Eq.~(\ref{nsrga}) gives 
\begin{equation}
n_{{\rm s}} \simeq 1-\frac{2(p+2)}{4N+p}\,,\qquad 
r\simeq\frac{16p}{4N+p}\simeq-8n_{{\rm t}}\,.
\label{nrstan}
\end{equation}
In Eq.~(\ref{nrstan}) we have not taken into 
account the contributions coming from the term $\mu$, 
because we do not have analytic an expression 
for general $p$.
Numerical calculations show that in the regime $B \ll 1$
both $n_{\rm s}$ and $r$ become smaller for $\mu>0$.
If $\mu<0$, then $n_{\rm s}$ get smaller, whereas
$r$ increases.

In the opposite limit where $B \gg 1$ it follows that 
\begin{equation}
\epsilon\simeq3\delta_{GX}\simeq\frac{p^{3/2}}{2B^{2}}x^{-(p+3)/2}
e^{-\mu x/2}\,,\qquad N\simeq\frac{B^{2}}{\sqrt{p}}\int_{x_{f}}^{x}x^{(p+1)/2}
e^{\mu x/2}dx\,,
\end{equation}
and $\delta_{X} \simeq px^{-(p+1)}e^{-\mu x}/(2B^4)$.
In order to have $N \approx 55$ for $B\gg1$ the integral inside the
expression of $N$ needs to be much smaller than 1, so that $x\ll1$
for $|\mu|={\cal O}(1)$. Using the approximation $|\mu x| \ll 1$, we have
$x_{f}^{(p+3)/2}\simeq p^{3/2}/(2B^{2})$ and 
\begin{equation}
N \simeq \frac{B^2}{\sqrt{p}} \frac{2}{p+3} x^{(p+3)/2}
\left[ 1+\frac{p+3}{2(p+5)}\mu x \right]-\frac{p}{p+3}\,.
\end{equation}
{}From Eqs.~(\ref{nsrga0}) and (\ref{nsgali}) it follows that 
\begin{eqnarray}
n_{{\rm s}} &\simeq&
1-\frac{3(p+1)}{(p+3)N+p}
\left[1-\frac{2(p-1)}{3(p+1)(p+5)} \mu x \right]\,, 
\label{nrgalie0} \\
r &\simeq& \frac{64\sqrt{6}}{9}\frac{p}{(p+3)N+p}
\left( 1-\frac{\mu x}{p+5} \right)\,.
\label{nrgali}
\end{eqnarray}
For $N=55$, in the limit where $\mu\to0$, one has $n_{{\rm s}}=0.9675$,
$r=0.1258$ for $p=2$ and $n_{{\rm s}}=0.9614$, $r=0.1791$ for
$p=4$. In the regime $B\gg1$ the current observations can be consistent
with both models. 
In the presence of the exponential coupling with positive $\mu$
the scalar spectral index gets larger for $p>1$, while 
the tensor-to-scalar ratio is smaller.

In the intermediate regime between $B \ll 1$ and $B \gg 1$ we evaluate
$n_{{\rm s}}$ and $r$ as follows. For given values of $p$, $\mu$,
and $B$ we identify the field value $x=\phi/\Mpl$ corresponding
to $N=55$ by integrating Eq.~(\ref{efoldGa}) numerically. We derive
$\delta_{X}$ and $\delta_{GX}$ from Eq.~(\ref{delXGX}) which allows us to
obtain $n_{{\rm s}}$ and $r$ by using the formulas (\ref{nscom})
and (\ref{rcom}). We have also solved the background equations numerically
to the end of inflation and confirmed that the above method provides
accurate estimation for $n_{{\rm s}}$ and $r$.

The theoretical values of $n_{{\rm s}}$ and $r$ for $\mu=1$ are
plotted in Figs.~\ref{gap2} and \ref{gap4} (corresponding to $p=2$
and $p=4$, respectively) with several different values of $B$. If
we choose larger $B$ starting from $B=0$, $n_{{\rm s}}$ decreases
up to some value of $B$, starts to increase, and finally decreases towards
the point given by Eq.~(\ref{nrgalie0}).
Meanwhile $r$ decreases up to some value of $B$ with 
a minimum smaller than 0.1, before starting to increase towards
the asymptotic value (\ref{nrgali}).
The above peculiar curved trajectories in the $(n_{\rm s}, r)$ plane
occur because of the presence of the exponential Galileon coupling 
with $\mu>0$.
For $\mu=0$ the theoretical curve can be approximated by a line 
that connects the two asymptotic points corresponding to 
$B \to 0$ and $B \to \infty$, see Figs.~\ref{gap2} and \ref{gap4}.

If $\mu<0$ and $B$ is increasing, then
$r$ increases up to some value of $B$, whereas 
$n_{\rm s}$ decreases. 
The maximum values of $r$ for $\mu=-1$
are about 0.35 and 0.68 for $p=2$ and $p=4$, respectively.
If $B$ is increased further, $r$ starts to decrease 
towards the point given by Eq.~(\ref{nrgali}) (with $n_{\rm s}$ starting
to increase at some value of $B$). 
Compared to the case $\mu>0$ this behaviour is not desirable
to satisfy the observational bounds, especially for $p=4$.
In the following discussion we shall therefore focus on the case of the 
positive $\mu$.

{}From Eqs.~(\ref{rcom}) and (\ref{ntcom}) one has 
\begin{equation}
\frac{r}{n_{{\rm t}}}=-8\frac{(1+4R_{G})^{3/2}}
{(1+6R_{G})^{1/2}(1+3R_{G})}\,,
\end{equation}
 where $R_{G}\equiv\delta_{GX}/\delta_{X}$. For $0\le R_{G}<\infty$
the ratio $r/n_{{\rm t}}$ is constrained to be in the narrow range
$-8.71<r/n_{{\rm t}}\le-8$. We carry out the CMB likelihood analysis
in terms of $n_{{\rm s}}$ and $r$ by using the two consistency relations
$r=-8n_{{\rm t}}$ and $r=-8.71n_{{\rm t}}$. We find that the observational
constraints on $n_{{\rm s}}$ and $r$ are similar in both cases.
Hence the constraints using the standard consistency relation $r=-8n_{{\rm t}}$
should be trustable even in the intermediate regime. Figure \ref{gap2}
shows that the quadratic inflaton potential is consistent with observations
even in the presence of the exponential Galileon coupling with $\mu=1$.
From Fig.~\ref{gap4} we find that the self coupling inflaton potential
can also be saved by taking into account the exponential Galileon coupling.

%%%%%%%%%%%%%%%%%%%%%%%%%%%%%
\begin{figure}
\begin{centering}
\includegraphics[width=3.2in,height=3.0in]{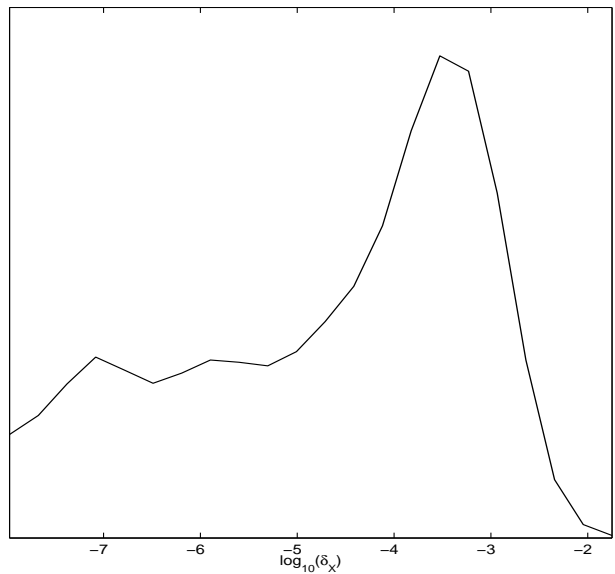} 
\par\end{centering}
\caption{1-dimensional marginalized probability distribution 
of the parameter $\delta_X$ for G-inflation with the
quartic potential $V(\phi)=\lambda\phi^{4}/4$ and $\mu=1$.
We use the fitting function (\ref{appeneq}) that gives the 
relation between $\delta_X$ and $\delta_{GX}$ 
for $N=55$ in the regime 
$10^{-8}<\delta_X<0.018$.
The parameter $\delta_X$ is constrained by the joint data analysis of 
WMAP 7 yr, LSS (including BAO), HST, SN Ia, and
BBN with the pivot scale $k_{0}=0.002$ Mpc$^{-1}$.}
\centering{}\label{constraintga} 
\end{figure}
%%%%%%%%%%%%%%%%%%%%%%%%%%%%

For the theoretical points shown in Figs.~\ref{gap2} and \ref{gap4}
we can evaluate the values of $\delta_X$ and $\delta_{GX}$ 
corresponding to $N=55$. It is then possible to derive fitting 
functions that relate $\delta_{GX}$ with $\delta_X$.
The fitting function for $p=4$ and $\mu=1$ 
is given in Eq.~(\ref{appeneq}) in the Appendix.
This allows us to run the CosmoMC code in terms of 
one inflationary parameter $\delta_X$.
In Fig.~\ref{constraintga} we show the 1-dimensional marginalized
probability distribution for $p=4$ and $\mu=1$ constrained by 
the joint data analysis of WMAP 7 yr, LSS, HST, SN Ia, and BBN. 
In the absence of the Galileon coupling ($\delta_{GX}=0$) one has 
$\delta_{X}=p/(4N+p)\simeq 0.018$
for $N=55$, which is observationally excluded.
In the opposite limit of large Galileon coupling such that $\delta_{GX}\gg\delta_{X}$,
it follows that $\delta_{GX}\simeq p/\{3[(p+3)N+p]\}=3.4\times10^{-3}$
for $N=55$. Since this case is marginally inside the 2$\sigma$ observational 
contour in Fig.~\ref{gap4}, we find a suppressed probability distribution
for smaller $\delta_X$ in Fig.~\ref{constraintga}.
The intermediate regime such as 
$10^{-4} \lesssim \delta_X \lesssim 10^{-3}$ 
is most favored observationally, because the corresponding 
theoretical points can be deep inside the 2$\sigma$ bound in Fig.~\ref{gap4}.
In Fig.~\ref{gap4} the theoretical point for $B=10^{-3/2}$
gives $\delta_X=3.5 \times 10^{-4}$, which actually corresponds to 
the highest probability in Fig.~\ref{constraintga}.
Hence the effect of the exponential Galileon coupling works to save
the self-coupling inflaton potential.

The scalar spectrum ${\cal P}_{\rm s}$ at the scale $k=0.002$ Mpc$^{-1}$ 
(for $n=1$) is subject to the WMAP normalization: 
\begin{equation}
{\cal P}_{\rm s}=\frac{\sqrt{3}}{\pi^{2}}\left(\frac{\Mpl}{M}\right)^{6}
\left(\frac{V_{0}}{\Mpl^{4}}\right)^{3}\frac{y^{1/2}x^{3p}
e^{2\mu x}}{(y-1)^{2}(2y+1)^{3/2}}\simeq2.4\times10^{-9}\,,
\label{WMAPnor}
\end{equation}
where $y\equiv(1+4pB^{4}x^{p-1}e^{\mu x})^{1/2}$. In the limit that
$B\gg1$ we obtain $m\approx10^{16}(10^{12}\,{\rm GeV}/M)$\,GeV
for $p=2$ (with $V_{0}=m^{2}\Mpl^{2}/2$) and 
$\lambda\approx(10^{12}\,{\rm GeV}/M)^{4}$
for $p=4$ (with $V_{0}=\lambda\Mpl^{4}/4$), which agree with those
obtained in Ref.~\cite{Kamada-etal-10} for $\mu=0$. In order to
have $B\gg1$ for $p=4$, we require that $\lambda\gg(M/\Mpl)^{3}$.
Combining this with the WMAP normalization, the mass scale $M$ is
constrained to be $M\ll10^{-4}M_{{\rm pl}}$. If we demand that the
coupling $\lambda$ is smaller than 1, this gives another constraint
$M>4\times10^{-7}\Mpl$. In the intermediate regime between $B\gg1$
and $B\ll1$ we need to solve Eq.~(\ref{WMAPnor}) to relate $M$
and $V_{0}$ after identifying the values of $x$ at $N=55$ numerically.
In the regime $B\ll1$ we recover the standard mass scales of inflaton:
$m/\Mpl \simeq 6.8 \times 10^{-6}$ for $p=2$ and 
$\lambda \simeq 2.0 \times 10^{-13}$ for $p=4$.

Since we regard $M$ to be a cutoff scale for the function $G(\phi,X)$,
the effective theory can be trusted as long as $H\lesssim M$. This
relation yields the constraint $B^{4}x^{p}\lesssim\Mpl/M$. For the
case \textbf{$B\gg1$} and $p=2$, we find that the effective theory
can be trusted for $x \lesssim M/m \approx (M/10^{14}\,{\rm GeV})^2$.
For $B\gg1$ and $p=4$, this constraint reduces to 
$x \lesssim \lambda^{-1/4} (M/\Mpl)^{1/2} \approx (M/10^{14}\,{\rm GeV})^{3/2}$.

Finally we note that the scalar propagation speed squared (\ref{scalarga}),
in the regime $\delta_{GX}\gg\delta_{X}$, reduces to $c_{s}^{2}\simeq2/(3n)$. Also, although the non-Gaussianity parameter $f_{{\rm NL}}^{{\rm equil}}$
is constrained to be small for $n=1$, it is possible to have 
$|f_{{\rm NL}}^{{\rm equil}}|\gg1$
for $n\gg1$. It would be of interest to see whether or not such models
can be compatible with observations.

%%%%%%%%%%%%%%%%%%%%%%
\section{Conclusions}
\label{conclusions} 
%%%%%%%%%%%%%%%%%%%%%%

In this paper we have studied the observational signatures 
of chaotic inflationary models with 
the potential $V(\phi)=V_0 (\phi/\Mpl)^p$ in the context of
modified gravitational theories.
In Einstein gravity the self-coupling potential $V(\phi)=\lambda \phi^4/4$
is excluded by CMB temperature anisotropy data, while 
the quadratic potential $V(\phi)=m^2 \phi^2/2$ is within the 
$2\sigma$ observational contour.
Our main aim here has been to clarify how various field couplings 
present in low-energy effective string theory modify the 
scalar/tensor power spectra generated during inflation.

We have found a number of new results summarized below.
\begin{itemize} 
\item (i) The inclusion of a non-canonical kinetic term $\omega(\phi)X$ with the exponential 
coupling $\omega(\phi)=e^{\mu \phi/\Mpl}$ ($\mu>0$) allows the 
chaotic inflation models that are in tension with observations to be made 
compatible with them.
We have studied the effects of the non-canonical kinetic coupling as well as the nonminimal coupling 
on the inflationary observables and have placed bounds on the strength 
of the couplings by using recent data from WMAP 7yr, BAO, and HST.

\item (ii) In Brans-Dicke theory we have found that the field potential of 
the form $V(\phi)=V_0 (\phi-\Mpl)^p$, where $p$ is close to 2, 
can be viable for inflation followed by a successful reheating.
We have evaluated the scalar index $n_{\rm s}$ and the tensor-to-scalar
ratio $r$ for the potential $V(\phi)=V_0 (\phi-\Mpl)^2$ and have shown
that $r$ decreases for smaller values of the BD parameter $\omega_{\rm BD}$. 
The models where $\omega_{\rm BD}$ is around the order of unity, 
which includes the $f(R)=R+R^2/(6M^2)$ model, is well within
the $1\sigma$ observational bound constrained by WMAP 7yr, BAO, and HST.

\item (iii) In the presence of the Gauss-Bonnet coupling of the form 
$\xi(\phi){\cal G}$, where $\xi(\phi)=\xi_0 e^{\mu \phi/\Mpl}$, 
we have found that the GB coupling with positive $\mu$ does not save 
the self-coupling potential $V(\phi)=\lambda \phi^4/4$.
For the quadratic potential $V(\phi)=m^2 \phi^2/2$ 
we have shown that the GB coupling needs to be suppressed 
($|\delta_{\xi}/\epsilon_s|<0.1$) from the CMB likelihood analysis.
If $\mu$ is negative then it is possible to lead to the decrease of 
both $|n_{\rm s}-1|$ and $r$ for negative $\xi_0$, 
but we require a large coupling constant, such as $|\xi_0| \sim 10^{10}$,
in order to produce a sizable effect on the inflationary observables.

\item (iv) In the presence of the Galileon-like self-interaction 
$G(\phi,X) \square \phi$, where $G(\phi,X)=\Phi (\phi) X^n$ 
and $\Phi \propto e^{\mu \phi/\Mpl}$, we have expressed the 
three inflationary observables $n_{\rm s}$, $r$, and $n_{\rm t}$
in terms of two slow-roll parameters $\delta_X$ and 
$\delta_{GX}$. 
In the regime where the Galileon term dominates over the standard 
kinetic term ($\delta_{GX} \gg \delta_X$) we have derived analytic
formulas for $n_{\rm s}$ and $r$ in terms of the number of 
e-foldings $N$, which recover the results obtained 
for $\mu=0$. We have shown that, for $\mu>0$, the Galileon 
term can lead to the compatibility of the chaotic inflationary 
potentials with current observations.
We have confirmed this property for the potential 
$V(\phi)=\lambda \phi^4/4$ by carrying out 
the CMB likelihood analysis.

\end{itemize} 

In summary, we have undertaken a unified study of the effects of 
a number of generalisations to the standard inflationary picture as motivated by low-energy effective 
string theory. We have found that a number of chaotic inflationary models
which are in tension with observations can be made compatible with them through
the addition of such terms. The stronger constraints on 
$n_{{\rm s}}$ and $r$ expected from the PLANCK satellite will provide an opportunity 
to further test the viability of such scenarios.

%%%%%%%%%%%%%%%%%%%%%%%%%%%%%%%%%%%%%
\section*{ACKNOWLEDGEMENTS}
\label{acknow} The work of A.\,D.\,F.\ and S.\,T.\ were supported
by the Grant-in-Aid for Scientific Research Fund of the JSPS Nos.~10271
and 30318802. S.\,T.\ also thanks financial support for the Grant-in-Aid
for Scientific Research on Innovative Areas (No.~21111006). 
J.\,E. was supported by a STFC studentship. This work has benefited from 
exchange visits supported by a JSPS and Royal Society bilateral grant.
%%%%%%%%%%%%%%%%%%%%%%%%%%%%%%%%%%%%%

\appendix

\section{Fitting function for G-inflation with an exponential coupling}
\label{appe}

\begin{figure}
\begin{centering}
\includegraphics[width=4.6in]{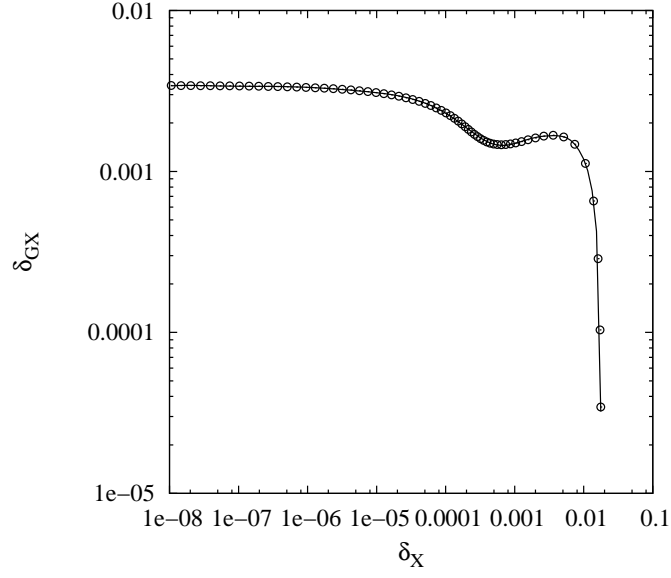} 
\par\end{centering}
\caption{Numerical data points corresponding to the values of $\delta_{GX}$ 
and $\delta_X$ satisfying the constraint $N=55$. 
Each data point corresponds to a particular value $B=10^{i/8}$ 
with $i=-32,\dots,32$.
The fitting function used for the CMB likelihood analysis is also plotted.}
\centering{}\label{fittof} 
\end{figure}

In this Appendix we present the fitting function for the quartic 
potential $V(\phi)=\lambda \phi^4/4$ in the presence of 
the Galileon-type coupling $G=-(1/M^{3})e^{\mu\phi/\Mpl}X$
with $\mu=1$. Numerically we find the field value $\phi$ giving 
$N=55$ before the end of inflation and 
evaluate $\delta_X$ and $\delta_{GX}$
for several different values of $B$ ($B=10^{i/8}$ with $i=-32,\dots,32$).
These slow-roll parameters can be approximated by the following 
fitting function (found by using the method of least squares)
\begin{eqnarray}
\delta_{GX} & = & -5.25192634579698+
540.210808997015\,\delta_{{X}}^{1/2}
-3509.55978587371\,\delta_{{X}}^{1/3}
+15290.159752272\,\delta_{{X}}^{1/4}\nonumber\\
&&-38509.9526724544\,\delta_{{X}}^{1/5}
+53949.4042466374\,\delta_{{X}}^{1/6}
-38908.1718682253\,\delta_{{X}}^{1/7}\nonumber \\
 &  & +11232.4296410833\,\delta_{{X}}^{1/8}
-224.28358682764\,\delta_{{X}}
+6155.30047533836\,{\delta_{{X}}^{2}}
\nonumber \\
& &-519243.629001884\,{\delta_{{X}}^{3}}
+38227861.764318\,{\delta_{{X}}^{4}}
-1897688289.07932\,{\delta_{{X}}^{5}}
+54200448383.7942\,{\delta_{{X}}^{6}}\nonumber \\
&&-665839723646.196\,{\delta_{{X}}^{7}} \,.
\label{appeneq}
\end{eqnarray}
We have used this expression in the regime $10^{-8}<\delta_X<0.018$
for the CMB likelihood analysis 
in Fig.~\ref{constraintga}. Finally, in Fig.~\ref{fittof}, we show both 
the numerical data and the fitting function $\delta_{GX}=\delta_{GX}(\delta_X)$. 
Since its inverse function, on the whole interval, is multivalued, we have used 
$\delta_X$ as the independent slow-roll parameter.

%%%%%%%%%%%%%%%%%%%%%%%%%%%%%%%%%%
%%%%%%%%%%%%%%%%%%%%%%%%%%%%%%%%%%

%%%%%%%%%%%%%%%%%%%%%


\begin{thebibliography}{References}
\bibitem{Star80} 
A.~A.~Starobinsky, 
%``A New Type Of Isotropic Cosmological
%Models Without Singularity,''
Phys.\ Lett.\ B \textbf{91}, 99 (1980).

\bibitem{infori} 
D.~Kazanas, 
%``Dynamics Of The Universe And
%Spontaneous Symmetry Breaking,''
Astrophys.\ J.\ \textbf{241} L59 (1980); 
K.~Sato, Mon.\ Not.\ R.\ Astron.\ Soc.
\textbf{195}, 467 (1981); 
A.~H.~Guth, %``The Inflationary Universe: A Possible Solution
%To The Horizon And Flatness Problems,''
Phys.\ Rev.\ D \textbf{23}, 347 (1981).

\bibitem{infper} 
V.~F.~Mukhanov and G.~V.~Chibisov, 
%``Quantum Fluctuation And 'Nonsingular' Universe.''
JETP Lett.\ \textbf{33}, 532 (1981); 
A.~H.~Guth and S.~Y.~Pi,
%``Fluctuations In The New Inflationary Universe,''
Phys.\ Rev.\ Lett.\ \textbf{49} (1982) 1110; 
S.~W.~Hawking, Phys.\ Lett.\ B
\textbf{115}, 295 (1982); 
A.~A.~Starobinsky, 
%``Dynamics Of Phase Transition In The New Inflationary
%Universe Scenario And Generation Of Perturbations,''
Phys.\ Lett.\ B \textbf{117} (1982) 175.

\bibitem{COBE} 
G.~F.~Smoot \textit{et al.}, 
%``Structure in the COBE differential microwave radiometer first year maps,''
Astrophys.\ J.\ \textbf{396}, L1-L5 (1992).

\bibitem{WMAP1} 
D.~N.~Spergel \textit{et al.} {[}WMAP Collaboration{]},
%``First year Wilkinson Microwave Anisotropy Probe (WMAP) observations:
%Determination of cosmological parameters,''
Astrophys.\ J.\ Suppl.\ \textbf{148}, 175 (2003).

\bibitem{WMAP7} 
E.~Komatsu \textit{et al.} {[}WMAP Collaboration{]},
%``Seven-Year Wilkinson Microwave Anisotropy Probe (WMAP) Observations:
%Cosmological Interpretation,''
Astrophys.\ J.\ Suppl.\ \textbf{192}, 18 (2011).

\bibitem{Tegmark} M.~Tegmark \textit{et al.} {[}SDSS Collaboration{]},
%``Cosmological parameters from SDSS and WMAP,''
Phys.\ Rev.\ D \textbf{69}, 103501 (2004).

\bibitem{Reid} 
B.~A.~Reid \textit{et al.}, 
%``Cosmological Constraints from the Clustering of the Sloan Digital Sky
%Survey DR7 Luminous Red Galaxies,''
Mon.\ Not.\ Roy.\ Astron.\ Soc.\ \textbf{404}, 60 (2010).

\bibitem{review} 
J.~E.~Lidsey \textit{et al.}, 
%``Reconstructing the inflaton potential: An overview,''
Rev.\ Mod.\ Phys.\ \textbf{69}, 373 (1997); 
D.~H.~Lyth and A.~Riotto,
%``Particle physics models of inflation and the cosmological density
%perturbation,''
Phys.\ Rept.\ \textbf{314}, 1 (1999); 
A.~D.~Linde, %``Particle physics and inflationary cosmology,''
arXiv:hep-th/0503203; 
B.~A.~Bassett, S.~Tsujikawa and D.~Wands,
%``Inflation dynamics and reheating,''
Rev.\ Mod.\ Phys.\ \textbf{78}, 537 (2006).

\bibitem{Linde-83} 
A.~D.~Linde, 
%  ``Chaotic Inflation,''
Phys.\ Lett.\ B \textbf{129}, 177 (1983).

\bibitem{Sotiriou} 
T.~P.~Sotiriou and V.~Faraoni, 
%``f(R) Theories Of Gravity,''
Rev.\ Mod.\ Phys.\ \textbf{82}, 451 (2010).

\bibitem{DT10}
A.~De Felice and S.~Tsujikawa, 
%``f(R) theories,''
Living Rev.\ Rel.\ \textbf{13}, 3 (2010).

\bibitem{supergravity} 
A.~B.~Goncharov and A.~D.~Linde, 
%``CHAOTIC INFLATION IN SUPERGRAVITY,''
Phys.\ Lett.\ B \textbf{139}, 27 (1984); 
M.~Kawasaki, M.~Yamaguchi and T.~Yanagida, 
%  ``Natural chaotic inflation in supergravity,''
Phys.\ Rev.\ Lett.\ \textbf{85}, 3572 (2000); 
P.~Brax and J.~Martin,
%``Shift symmetry and inflation in supergravity,''
Phys.\ Rev.\ D \textbf{72}, 023518 (2005); 
M.~B.~Einhorn and D.~R.~T.~Jones,
%``Inflation with Non-minimal Gravitational Couplings in Supergravity,''
JHEP \textbf{1003}, 026 (2010); 
S.~Ferrara, R.~Kallosh, A.~Linde, A.~Marrani and A.~Van Proeyen, 
%``Jordan Frame Supergravity and Inflation in NMSSM,''
Phys. Rev. \textbf{D82}, 045003 (2010); 
H.~M.~Lee, 
%``Chaotic inflation in Jordan frame supergravity,''
JCAP \textbf{1008}, 003 (2010); 
R.~Kallosh and A.~Linde, 
%``New models of chaotic inflation in supergravity,''
JCAP \textbf{1011}, 011 (2010); 
R.~Kallosh, A.~Linde and T.~Rube,
%``General inflaton potentials in supergravity,''
Phys.\ Rev.\ D \textbf{83}, 043507 (2011).

\bibitem{superstring} 
E.~Silverstein and A.~Westphal, 
%``Monodromy in the CMB: Gravity Waves and String Inflation,''
Phys.\ Rev.\ D \textbf{78}, 106003 (2008); 
L.~McAllister, E.~Silverstein and A.~Westphal, 
%``Gravity Waves and Linear Inflation from Axion Monodromy,''
Phys.\ Rev.\ D \textbf{82}, 046003 (2010); R.~Flauger, L.~McAllister,
E.~Pajer, A.~Westphal and G.~Xu, 
%``Oscillations in the CMB from Axion Monodromy Inflation,''
JCAP \textbf{1006}, 009 (2010); 
M.~Berg, E.~Pajer and S.~Sjors,
%  ``Dante's Inferno,''
Phys.\ Rev.\ \textbf{D81}, 103535 (2010); 
X.~Dong, B.~Horn, E.~Silverstein and A.~Westphal, 
%``Simple exercises to flatten your potential,''
arXiv:1011.4521 {[}hep-th{]}.

\bibitem{Futamase} 
T.~Futamase and K.~i.~Maeda, 
%``CHAOTIC INFLATIONARY SCENARIO IN MODELS HAVING NONMINIMAL COUPLING WITH
%CURVATURE,''
Phys.\ Rev.\ D \textbf{39}, 399 (1989).

\bibitem{Fakir} 
R.~Fakir and W.~G.~Unruh, 
%``Improvement on cosmological chaotic inflation through nonminimal
%coupling,''
Phys.\ Rev.\ D \textbf{41}, 1783 (1990).

\bibitem{Bez} 
F.~L.~Bezrukov and M.~Shaposhnikov, 
%``The Standard Model Higgs boson as the inflaton,''
Phys.\ Lett.\ B \textbf{659}, 703 (2008).

\bibitem{GBearly} 
I.~Antoniadis, J.~Rizos and K.~Tamvakis, Nucl.\ Phys.\ B
\textbf{415}, 497 (1994); 
R.~Easther and K.~i.~Maeda, 
%``One-Loop Superstring Cosmology and 
%the Non-Singular Universe,''
Phys.\ Rev.\ D \textbf{54}, 7252 (1996); 
S.~Kawai and J.~Soda,
%``Non-singular Bianchi type I cosmological solutions from 
%1-loop  superstring effective action,''
Phys.\ Rev.\ D \textbf{59}, 063506 (1999).

\bibitem{Armendariz} 
C.~Armendariz-Picon, T.~Damour and V.~F.~Mukhanov,
%``k-Inflation,''
 Phys.\ Lett.\ B \textbf{458}, 209 (1999).

\bibitem{Ohta} 
K.~i.~Maeda and N.~Ohta, 
%``Inflation from M-theory with fourth-order corrections 
%and large extra dimensions,''
Phys.\ Lett.\ B \textbf{597}, 400 (2004); Phys.\ Rev.\ D \textbf{71},
063520 (2005); Z.~K.~Guo, N.~Ohta and S.~Tsujikawa, 
%``Realizing scale-invariant density perturbations 
%in low-energy effective string theory,''
Phys.\ Rev.\ D \textbf{75}, 023520 (2007).

\bibitem{Satoh} 
M.~Satoh and J.~Soda, 
%``Higher Curvature Corrections to Primordial Fluctuations in Slow-roll Inflation,''
JCAP \textbf{0809}, 019 (2008).

\bibitem{Nicolis} A.~Nicolis, R.~Rattazzi and E.~Trincherini,
%``The galileon as a local modification of gravity,''
Phys.\ Rev.\ D \textbf{79}, 064036 (2009).

\bibitem{Cremi} 
P.~Creminelli, A.~Nicolis and E.~Trincherini,
%``Galilean Genesis: an alternative to inflation,''
JCAP \textbf{1011}, 021 (2010); 
C.~Burrage, C.~de Rham, D.~Seery and A.~J.~Tolley, 
%``Galileon inflation,''
JCAP \textbf{1101}, 014 (2011); 
P.~Creminelli \textit{et al.,} 
%``Galilean symmetry in the effective theory of inflation: new shapes of
%non-Gaussianity,''
JCAP \textbf{1102}, 006 (2011);
A.~Naruko and M.~Sasaki, 
%``Conservation of the nonlinear curvature perturbation in generic
%single-field inflation,''
Class.\ Quant.\ Grav.\ \textbf{28}, 072001 (2011); 
C.~Burrage, C.~de Rham and L.~Heisenberg, 
%``de Sitter Galileon,''
arXiv:1104.0155 {[}hep-th{]}.

\bibitem{Kobayashi-etal-10} 
T.~Kobayashi, M.~Yamaguchi and J.~Yokoyama,
%``G-inflation: inflation driven by the Galileon field,''
Phys.\ Rev.\ Lett.\ \textbf{105}, 231302 (2010).

\bibitem{nonminimalper} 
N.~Makino and M.~Sasaki, 
%``The Density perturbation in the chaotic inflation with nonminimal coupling,''
Prog.\ Theor.\ Phys.\ \textbf{86}, 103 (1991); 
R.~Fakir, S.~Habib and W.~Unruh, 
%``Cosmological density perturbations with modified gravity,''
Astrophys.\ J.\ \textbf{394}, 396 (1992); 
D.~I.~Kaiser, 
%``Primordial spectral indices from generalized Einstein theories,''
Phys.\ Rev.\ D \textbf{52}, 4295 (1995); J.~c.~Hwang and H.~Noh,
%``Cosmological perturbations in generalized 
%gravity theories,''
Class.\ Quant.\ Grav. \textbf{14}, 3327 (1998); 
E.~Komatsu and T.~Futamase, 
%``Constraints on the chaotic inflationary scenario with a nonminimally
%coupled 'inflaton' field from the cosmic microwave background radiation
%anisotropy,''
Phys.\ Rev.\ D \textbf{58}, 023004 (1998).

\bibitem{Komatsuper} 
E.~Komatsu and T.~Futamase, 
%``Complete constraints on a nonminimally coupled chaotic inflationary
%scenario from the cosmic microwave background,''
Phys.\ Rev.\ D \textbf{59}, 064029 (1999).

\bibitem{Gumjudpai} 
S.~Tsujikawa and B.~Gumjudpai, 
%``Density perturbations in generalized Einstein scenarios and constraints on
%nonminimal couplings from the Cosmic Microwave Background,''
Phys.\ Rev.\ D \textbf{69}, 123523 (2004).

\bibitem{Linde} 
A.~Linde, M.~Noorbala and A.~Westphal, 
%``Observational consequences of chaotic inflation with nonminimal 
%coupling to gravity,''
JCAP \textbf{1103}, 013 (2011).

\bibitem{Higgspapers} 
A.~O.~Barvinsky, A.~Y.~.Kamenshchik and A.~A.~Starobinsky, 
%``Inflation scenario via the Standard Model Higgs boson and LHC,''
JCAP \textbf{0811}, 021 (2008); 
F.~L.~Bezrukov, A.~Magnin and M.~Shaposhnikov,
%``Standard Model Higgs boson mass from inflation,''
Phys.\ Lett.\ \textbf{B675}, 88-92 (2009); 
F.~Bezrukov and M.~Shaposhnikov,
%``Standard Model Higgs boson mass from inflation: Two loop analysis,''
JHEP \textbf{0907}, 089 (2009); 
A.~De Simone, M.~P.~Hertzberg and F.~Wilczek, 
%``Running Inflation in the Standard Model,''
Phys.\ Lett.\ \textbf{B678}, 1-8 (2009); 
A.~O.~Barvinsky \textit{et al.,} 
%``Asymptotic freedom in inflationary cosmology 
%with a non-minimally coupled Higgs field,''
JCAP \textbf{0912}, 003 (2009); 
D.~I.~Kaiser and A.~T.~Todhunter,
%``Primordial Perturbations from Multifield Inflation with 
%Nonminimal Couplings,''
Phys.\ Rev.\ D \textbf{81}, 124037 (2010).

\bibitem{Kamada-etal-10} 
K.~Kamada, T.~Kobayashi, M.~Yamaguchi
and J.~Yokoyama, 
%``Higgs G-inflation,''
Phys.\ Rev.\ D \textbf{83}, 083515 (2011).

\bibitem{Galipapers} 
C.~Deffayet, G.~Esposito-Farese and A.~Vikman,
%``Covariant Galileon,''
Phys.\ Rev.\  D {\bf 79}, 084003 (2009);
C.~Deffayet, S.~Deser and G.~Esposito-Farese,
%``Generalized Galileons: All scalar models whose curved 
%background extensions
%maintain second-order field equations and stress-tensors,''
Phys.\ Rev.\  D {\bf 80}, 064015 (2009);
R.~Gannouji and M.~Sami,
%``Galileon gravity and its relevance to late time cosmic acceleration,''
Phys.\ Rev.\  D {\bf 82}, 024011 (2010);
A.~De Felice and S.~Tsujikawa,
%``Cosmology of a covariant Galileon field,''
Phys.\ Rev.\ Lett.\  {\bf 105}, 111301 (2010);
arXiv:1008.4236 [hep-th];
S.~Nesseris, A.~De Felice and S.~Tsujikawa,
%``Observational constraints on Galileon cosmology,''
Phys.\ Rev.\  D {\bf 82}, 124054 (2010).

\bibitem{corrections} 
R.~Metsaev and A.~Tseytlin, Nucl.\ Phys.\ \textbf{B}
293, 385 (1987).

\bibitem{ADM} 
R.~L.~Arnowitt, S.~Deser and C.~W.~Misner, 
Phys. Rev. {\bf 117}, 1595-1602 (1960).

\bibitem{Maldacena} 
J.~M.~Maldacena,
%``Non-Gaussian features of primordial fluctuations in single field
%inflationary models,''
JHEP {\bf 0305}, 013 (2003).

\bibitem{DT11} 
A.~De Felice and S.~Tsujikawa, 
%``Primordial non-Gaussianities in general modified gravitational models of inflation,''
JCAP \textbf{1104}, 029 (2011).

\bibitem{Mizuno} 
S.~Mizuno and K.~Koyama, 
%``Primordial non-Gaussianity from the DBI Galileons,''
Phys.\ Rev.\ D \textbf{82}, 103518 (2010).

\bibitem{Kobayashi-etal-11} 
T.~Kobayashi, M.~Yamaguchi and J.~Yokoyama,
%``Primordial non-Gaussianity from G-inflation,''
 arXiv:1103.1740 {[}hep-th{]}.

\bibitem{Maeda}
K.~i.~Maeda,
%``Towards the Einstein-Hilbert Action via Conformal Transformation,''
Phys.\ Rev.\  D {\bf 39}, 3159 (1989).

\bibitem{Leach}
A.~R.~Liddle and S.~M.~Leach,
%``How long before the end of inflation were observable perturbations
%produced?,''
Phys.\ Rev.\  D {\bf 68}, 103503 (2003).

\bibitem{Catena} 
R.~Catena, M.~Pietroni and L.~Scarabello, 
%``Einstein and Jordan frames reconciled: a frame-invariant approach to
%scalar-tensor cosmology,''
Phys.\ Rev.\ D \textbf{76}, 084039 (2007); 
N.~Deruelle and M.~Sasaki,
%``Conformal equivalence in classical gravity: the example of 'veiled' General Relativity,''
arXiv:1007.3563 {[}gr-qc{]}.

\bibitem{BAO} 
W.~Percival \textit{et al.}, Mon.\ Not.\ Roy.\ Astron.\ Soc.\ 1741
(2009).

\bibitem{HST} 
A.~G.~Riess \textit{et al.}, 
%``A Redetermination of the Hubble Constant with the Hubble Space Telescope
%from a Differential Distance Ladder,''
Astrophys.\ J.\ \textbf{699}, 539 (2009).

\bibitem{Lambert} 
R.~Corless {\it et al.,}
Advances in Computational Mathematics, {\bf 5}, 329 (1996).

\bibitem{Brans} 
C.~Brans and R.~H.~Dicke, 
%``Mach's principle and a relativistic theory of gravitation,''
Phys.\ Rev.\ \textbf{124}, 925 (1961).

\bibitem{Ohashi} 
J.~Ohashi and S.~Tsujikawa, 
%``Observational constraints on assisted k-inflation,''
Phys.\ Rev.\ D \textbf{83}, 103522 (2011).

\bibitem{Ohanlon} 
J.~O'Hanlon, Phys.\ Rev.\ Lett.\ \textbf{29},
137 (1972); 
T.~Chiba, 
%``1/R gravity and scalar-tensor gravity,''
Phys.\ Lett.\ B \textbf{575}, 1 (2003).

\bibitem{fRspe}
A.~A.~Starobinsky, Pisma v Astron. Zh.\,9, 579
(1983) {[}Sov. Astron. Lett. 9, 302 (1983){]}; 
L.~A.~Kofman, V.~F.~Mukhanov and D.~Y.~Pogosian, 
%``EVOLUTION OF INHOMOGENEITIES IN INFLATIONARY MODELS IN A THEORY OF
%GRAVITATION WITH HIGHER DERIVATIVES,''
Sov.\ Phys.\ JETP \textbf{66}, 433 (1987) {[}Zh.\ Eksp.\ Teor.\ Fiz.\ \textbf{93},
769 (1987){]}; 
J.~c.~Hwang and H.~Noh, 
%``$f(R)$ gravity theory and CMBR constraints,''
Phys.\ Lett.\ B \textbf{506}, 13 (2001).

\bibitem{Contaldi} 
N.~Tamanini and C.~R.~Contaldi, 
%``Inflationary Perturbations in Palatini Generalised Gravity,''
Phys.\ Rev.\ D \textbf{83}, 044018 (2011).

\bibitem{Stewart} 
E.~D.~Stewart and D.~H.~Lyth, 
%``A more accurate analytic calculation of the spectrum of cosmological
%perturbations produced during inflation,''
Phys.\ Lett.\ B \textbf{302}, 171 (1993).

\bibitem{PLANCK} 
[PLANCK Collaboration], arXiv: astro-ph/0604069.

\bibitem{cosmomc} 
http://cosmologist.info/cosmomc/

\bibitem{SNIa} 
M.~Kowalski \textit{et al.} {[}Supernova Cosmology
Project Collaboration{]}, 
%``Improved Cosmological Constraints from New, Old and Combined Supernova
%Datasets,''
Astrophys.\ J.\ \textbf{686}, 749 (2008).

\bibitem{BBN} 
S.~Burles and D.~Tytler, 
%``The Deuterium abundance towards Q1937-1009,''
Astrophys.\ J.\ \textbf{499}, 699 (1998).


\end{thebibliography}
\end{document}